\documentclass[aps,prd,secnumarabic,nobibnotes,twocolumn]{revtex4-1}
\usepackage{amsfonts}
\usepackage{mathrsfs}
\usepackage{amsmath}
\usepackage{color}
\usepackage{natbib}
\usepackage{graphicx}
\usepackage{bm}
\usepackage{amssymb}
\usepackage{xspace}
\usepackage{epstopdf}
\usepackage{dcolumn}
\usepackage{multirow}
\usepackage[colorlinks=true, letterpaper=true, pdfstartview=FitV, linkcolor=blue, citecolor=blue, urlcolor=blue]{hyperref}
\usepackage{wrapfig}
\usepackage{array}

\makeatletter

\begin{document}
\title{Magnetic Real Chern Insulator in 2D Metal-Organic Frameworks}
\author{Xiaoming Zhang}
\altaffiliation{X. M. Zhang and T. L. He contribute equally to this work.}
\address{State Key Laboratory of Reliability and Intelligence of Electrical Equipment}
\address{School of Materials Science and Engineering, Hebei University of Technology, Tianjin 300130, China.}

\author{Tingli He}
\altaffiliation{X. M. Zhang and T. L. He contribute equally to this work.}
\address{State Key Laboratory of Reliability and Intelligence of Electrical Equipment}
\address{School of Materials Science and Engineering, Hebei University of Technology, Tianjin 300130, China.}

\author{Ying Liu}
\address{State Key Laboratory of Reliability and Intelligence of Electrical Equipment}
\address{School of Materials Science and Engineering, Hebei University of Technology, Tianjin 300130, China.}

\author{Xuefang Dai}
\address{State Key Laboratory of Reliability and Intelligence of Electrical Equipment}
\address{School of Materials Science and Engineering, Hebei University of Technology, Tianjin 300130, China.}

\author{Guodong Liu}\email{gdliu1978@126.com}
\address{State Key Laboratory of Reliability and Intelligence of Electrical Equipment}
\address{School of Materials Science and Engineering, Hebei University of Technology, Tianjin 300130, China.}

\author{Cong Chen}
\address{Department of Physics, The University of Hong Kong, Hong Kong, China}
\address{HKU-UCAS Joint Institute of Theoretical and Computational Physics at Hong Kong, China}

\author{Weikang Wu}
\address{Key Laboratory for Liquid-Solid Structural Evolution and Processing of Materials (Ministry of Education), Shandong University, Jinan 250061, China}

\author{Jiaojiao Zhu}
\address{Research Laboratory for Quantum Materials, Singapore University of Technology and Design, Singapore 487372, Singapore}

\author{Shengyuan A. Yang}
\address{Research Laboratory for Quantum Materials, Singapore University of Technology and Design, Singapore 487372, Singapore}

\begin{abstract}
Real Chern insulators have attracted great interest, but so far, their material realization is limited to
nonmagnetic crystals and to systems without spin-orbit coupling. Here, we reveal magnetic real Chern insulator (MRCI) state in a recently synthesized metal-organic framework material Co$_3$(HITP)$_2$. Its ground state with in-plane ferromagnetic ordering hosts a nontrivial real Chern number,
enabled by the $C_{2z}T$ symmetry and robust against spin-orbit coupling. Distinct from previous nonmagnetic examples,
the topological corner zero-modes of MRCI are spin-polarized. Furthermore, under small tensile strains, the material  undergoes a topological phase transition from MRCI to a magnetic double-Weyl semimetal phase, via a pseudospin-1 critical state. Similar physics can also be found in closely related materials Mn$_3$(HITP)$_2$ and Fe$_3$(HITP)$_2$, which are also existing. Possible experimental detections and implications of an emerging magnetic flat band in the system are discussed.

\end{abstract}
\maketitle

Materials with nontrivial band structure topology have been a focus in condensed matter research~\cite{RevModPhys.82.3045,RevModPhys.83.1057,shen2012topological,RevModPhys.88.021004,RevModPhys.90.015001}.
In most cases, such topologies are defined with respect to the occupied band eigenstates, which are complex functions. However, under certain symmetry constraint, e.g., in the presence of spacetime inversion symmetry $PT$ and without spin-orbit coupling (SOC), the band eigenstates are required to be \emph{real}~\cite{PhysRevLett.116.156402,PhysRevLett.118.056401}. The topological classification of such real band structures (a real vector bundle over the Brillouin zone torus) is fundamentally different from the complex cases~\cite{PhysRevLett.116.156402,PhysRevB.96.155105}.
This leads to a family of real topological phases, with unique physical properties, such as $\mathbb{Z}_2$-charged nodal surfaces~\cite{PhysRevB.97.115125}, nodal-loop linking structure~\cite{PhysRevLett.121.106403}, unconventional bulk-boundary correspondence~\cite{PhysRevLett.125.126403,PhysRevB.104.085205}, non-Abelian braiding~\cite{doi:10.1126/science.aau8740}, and etc. Notably, in two dimensions (2D), there exists a
real Chern insulator state, characterized by real Chern number $\nu_R$ (also known as the second Stiefel-Whitney number)~\cite{PhysRevLett.118.056401}, which generically hosts protected corner zero-modes and can exhibit boundary phase transitions~\cite{PhysRevLett.125.126403,PhysRevB.104.085205}. Recent studies revealed existing materials such as graphdiyne and graphyne as its candidates~\cite{PhysRevLett.123.256402,PhysRevB.104.085205,lee2020two,PhysRevLett.128.026405,PhysRevB.105.085123}, which attracted great interest on real Chern insulators.

A current trend in topological materials research is to extend the study from nonmagnetic to magnetic systems, aiming to explore spin-polarized topological modes and the interplay between band topology and magnetism. Can real Chern insulator appear also in a magnetic system? The answer is yes. This concept is not forbidden by time reversal symmetry ($T$) breaking.
However, long-range magnetic ordering usually requires non-negligible SOC, especially for low-dimensional systems.
Considering SOC, $PT$ can no longer enforce a real band structure~\footnote{
This is due to the algebra $(PT)^2=\pm 1$ for spinless/spinful cases. Nevertheless, with certain gauge flux distribution, the two cases may be switched. See Ref~\cite{PhysRevLett.126.196402}.
}, but for 2D, $C_{2z}T$ symmetry can still maintain the reality condition (here, $z$ is the direction normal to the 2D plane)~\cite{PhysRevB.99.235125,PhysRevX.9.021013}, so in principle there should exist $C_{2z}T$-symmetric magnetic real Chern insulators (MRCIs). Then, the next question is: Can we identify any concrete MRCI material? To our knowledge, this question has not been addressed. The lack of material candidates poses an outstanding challenge, which hinders experimental studies on this fascinating topological phase.

In this work, we reveal MRCI in an existing 2D metal-organic framework (MOF) material.  MOFs are an extensive class of crystalline materials consisting of metal ions or clusters connected by organic ligands~\cite{li1999design,WOS:000323652300033}. Current techniques can enable growth of MOFs with designed composition and geometry. By incorporating transition metal ions, robust magnetism can be achieved in 2D MOFs~\cite{WOS:000445129300013,WOS:000437677800014,https://doi.org/10.1002/smll.201804845}. Besides their traditional application in energy storage~\cite{C5EE00762C,B802256A,WOS:000606751900002}, catalysis~\cite{B807080F,C4CS00094C,https://doi.org/10.1002/adma.201703663}, and sensors~\cite{doi:10.1021/cr200324t,C9CS00778D}, recent studies also reported their potential to realize topological states, such as quantum spin Hall and quantum anomalous Hall insulators~\cite{PhysRevLett.110.196801,PhysRevB.94.081102,doi:10.1021/acs.nanolett.9b05242,doi:10.1021/acs.accounts.0c00652}.

Here, using first-principles calculations and theoretical analysis, we demonstrate the recently synthesized
2D MOF Co$_3$(HITP)$_2$ (HITP stands for the ligand 2,3,6,7,10,11-hexaiminotriphenylene)~\cite{doi:10.1021/jacs.0c04458,https://doi.org/10.1002/admt.202000941} as the first example of MRCI.
We show that monolayer Co$_3$(HITP)$_2$ has in-plane ferromagnetism (FM) in its ground state, which respects the required $C_{2z}T$ symmetry. The bulk bands form a MRCI with nontrivial real Chern number $\nu_R=1$, robust against SOC.
We confirm the resulting $C_{2z}T$-pairs of corner zero-modes. Distinct from nonmagnetic real Chern insulators like the graphynes, the corner zero-modes here are strongly spin-polarized, which can be detected by spin-resolved scanning tunneling spectroscopy (STS). In addition, we find that under a small strain ($\sim 1\%$), the band gap closes in one spin channel, forming a pseudospin-1 critical point. Further increasing strain transforms the system into
a magnetic double-Weyl semimetal phase, and there appears a singular flat band with very small bandwidth, which could be an ideal platform to explore various quantum and interaction effects. Furthermore, we show that similar physics also exist in Mn$_3$(HITP)$_2$ and Fe$_3$(HITP)$_2$, which have also been synthesized in experiment~\cite{https://doi.org/10.1002/admt.202000941,doi:10.1021/acs.jpcc.0c08140}.

\begin{figure}
\includegraphics[width=8.8cm]{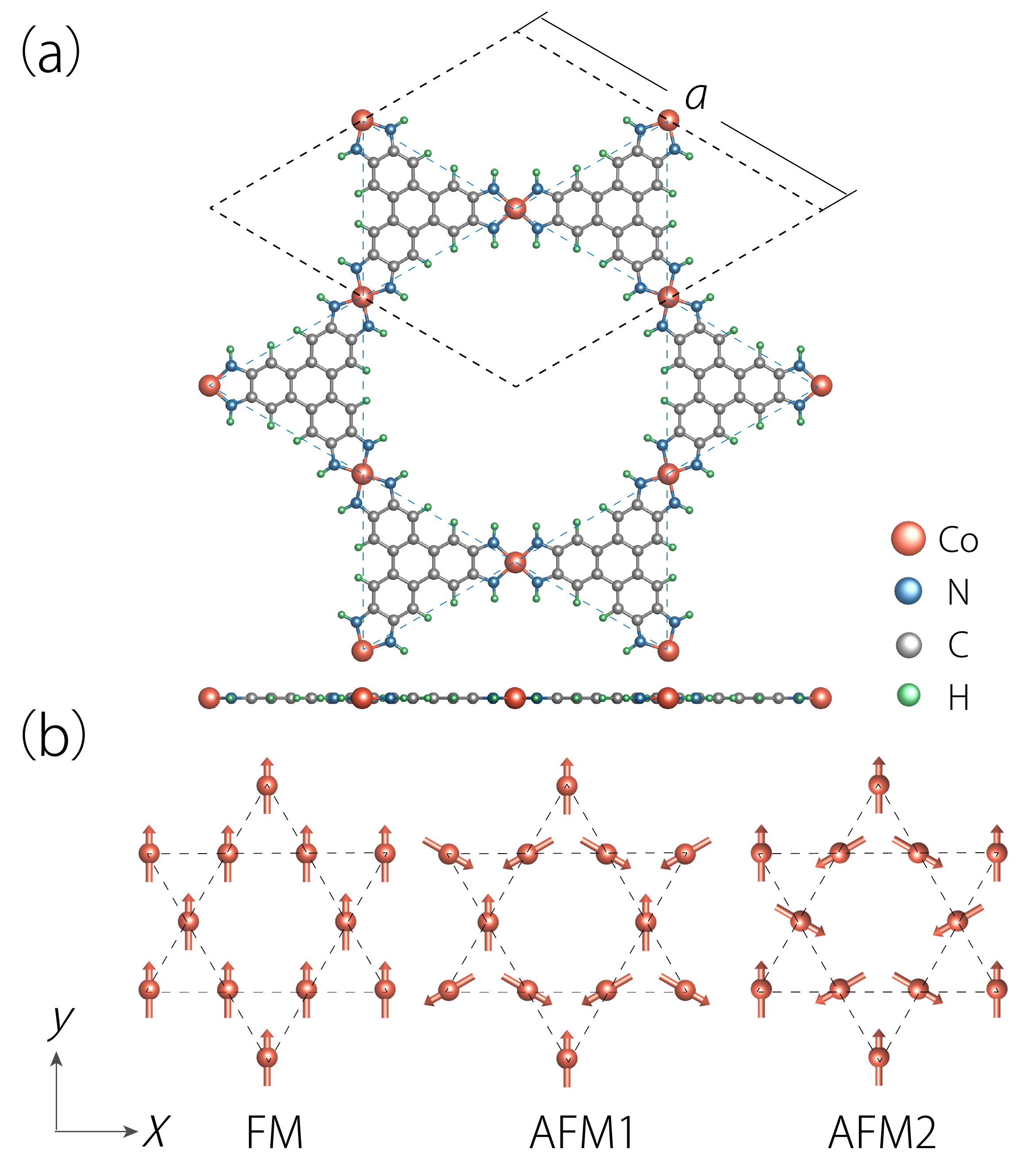}
\caption{(a) Top and side views of monolayer Co$_3$(HITP)$_2$. The dash box marks the unit cell. (b) Illustrations of typical magnetic configurations. The ground state has in-plane FM ordering. Two typical non-collinear antiferromagnetic configurations (AFM1 and AFM2) are also considered.
\label{fig1}}
\end{figure}

\emph{{\color{blue} Crystal structure and magnetic ordering.}} High-quality Co$_3$(HITP)$_2$ crystals have been synthesized in recent experiments~\cite{doi:10.1021/jacs.0c04458,https://doi.org/10.1002/admt.202000941}. It is a van der Waals layered material. Ultrathin Co$_3$(HITP)$_2$ 2D layers were demonstrated via exfoliation method~\cite{https://doi.org/10.1002/admt.202000941}. The crystal structure of monolayer Co$_3$(HITP)$_2$ is shown in Fig.~\ref{fig1}(a). It is a completely flat sheet with single-atom thickness. In the framework structure, Co ions form a kagome lattice, linked by HITP ligands. The crystal lattice has space group $P6/mmm$ (No.191), which preserves $C_{2z}$ (and also $P$) symmetry. The optimized lattice constant from our calculation (see Supplemental Material~\cite{refSM} for the calculation details) is $a= 21.93 $\AA$ $, consistent with previous works~\cite{doi:10.1021/acs.jpcc.0c01143,D2CP02612K}. Our calculation also shows that bulk Co$_3$(HITP)$_2$ has an extremely low exfoliation energy $\sim 0.07$ J/m$^2$, which agrees with the ease of exfoliation found in experiment~\cite{https://doi.org/10.1002/admt.202000941}.

Transition metal ions with partially filled $d$ shells are typical sources for magnetism. In Co$_3$(HITP)$_2$ monolayer, the magnetism is mainly from Co$^{2+}$ ions. Previous studies concluded that its ground state is FM with in-plane magnetization~\cite{D2CP02612K}. We confirm this result by comparing energies of different magnetic configurations, which also includes the typical non-collinear antiferromagnetic configurations in Fig.~\ref{fig1}(b)~\cite{refSM}. The obtained magnetic moment at Co site is $\sim 1.2 \mu_{B}$. Previous Monte Carlo simulations predicted that the FM ground state has a Curie temperature $\sim 11$ K~\cite{D2CP02612K}. It is important to note that the ground state magnetism preserves the combined $C_{2z}T$ symmetry, which is crucial for MRCI state.

\begin{figure}
\includegraphics[width=8.8cm]{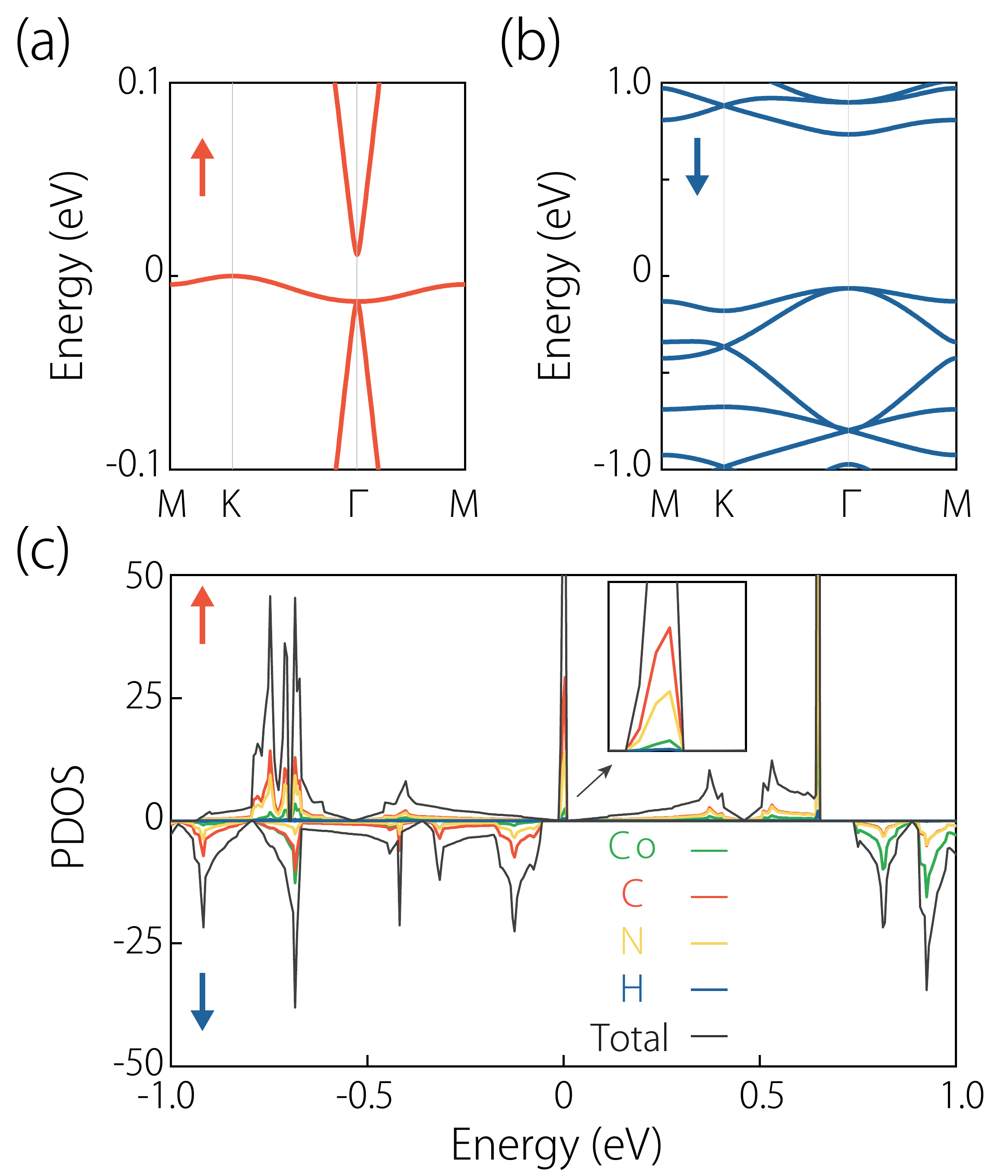}
\caption{Band structure of monolayer Co$_3$(HITP)$_2$ in the absence of SOC. (a) and (b) show the two spin channels. (c) shows projected density of states.
\label{fig2}}
\end{figure}

\emph{{\color{blue} MRCI state.}} Now, we examine the electronic band structure for monolayer Co$_3$(HITP)$_2$ in the FM ground state. First, we investigate the result in the absence of SOC, which is plotted in Fig.~\ref{fig2} separately for the two spin channels. One can see that the system is a narrow-gap semiconductor. One spin channel, which we denote as spin-down channel, has a relatively large band gap of 796 meV (Fig.~\ref{fig2}(b)). Meanwhile, the other channel, denoted as spin-up, has a much smaller gap $\sim 10.4$ meV (Fig.~\ref{fig2}(a)). From projected density of states (PDOS) in Fig.~\ref{fig2}(c), one can see that the low-energy bands are mainly from N and C orbitals. Especially, the states around band edges are spin-polarized in the spin-up channel.

In the absence of SOC, the two spin channels are decoupled. Each one can be regarded as an effective \emph{spinless} system, which preserves all the crystal symmetries (i.e., the original ones without magnetism) including $T$~\cite{vanderbilt2018berry}. In other words, the system consists of two spinless subsystems, and for \emph{each} subsystem (labeled by $\sigma=\uparrow,\downarrow$) we can define a real Chern number $\nu_R^\sigma$  since it preserves both $T$ and $C_{2z}$ symmetries and $(C_{2z}T)^2=1$. For such a case, there is a convenient way to evaluate $\nu_R^\sigma$ from $C_{2z}$ eigenvalues of the valence bands at the four $C_{2z}$-invariant momentum points $\Gamma_i$ ($i=1,\cdots,4$) in Brillouin zone.
Explicitly, for channel $\sigma$, we have~\cite{PhysRevLett.121.106403,PhysRevB.104.085205}
\begin{equation}\label{nu}
(-1)^{\nu_{R}^\sigma} = \prod_{i=1}^{4} (-1)^{\lfloor (n^{\sigma}_{i,-})/2 \rfloor}
\end{equation}
where  $\lfloor...\rfloor$ is the floor function, and $n^{\sigma}_{i,-}$ is the number of valence states
at $\Gamma_i$ which have negative $C_{2z}$ eigenvalues. These eigenvalues can be directly analyzed
from first-principles calculations. The result is shown in Table~\ref{tab1}.  Using formula (\ref{nu}), we find that the spin-up channel has a nontrivial $\nu_{R}^\uparrow=1$, whereas the spin-down channel is trivial with $\nu_{R}^\downarrow=0$. Therefore, the real Chern number for the whole system is given by $\nu^R=\nu_{R}^\uparrow+\nu_{R}^\downarrow=1$, i.e., the system represents a MRCI when SOC is not considered.

\begin{table}

\renewcommand\arraystretch{1.2}
\caption{$C_{2z}$ eigenvalues of valence states at the four $C_{2z}$-invariant momentum points.}
\label{tab1}       
\begin{center}
\setlength{\tabcolsep}{5.65mm}{
\begin{tabular}{cccccc}
\hline
 \hline & \multicolumn{2}{c}{spin-up} & & \multicolumn{2}{c}{spin-down} \\
 \cline { 2 - 3 } \cline { 5 - 6} & $\Gamma$ & $M$  & & $\Gamma$ & $M$  \\
 \hline$n_{+}$ & 66 & 64  & & 66 & 62  \\$n_{-}$ & 63 & 65  & & 60 & 64 \\
$\nu_{R}$  & \multicolumn{2}{c}{1} & & \multicolumn{2}{c}{0} \\
 \hline
 \hline
\end{tabular}}
\end{center}
\end{table}

\begin{figure}
\includegraphics[width=8.8cm]{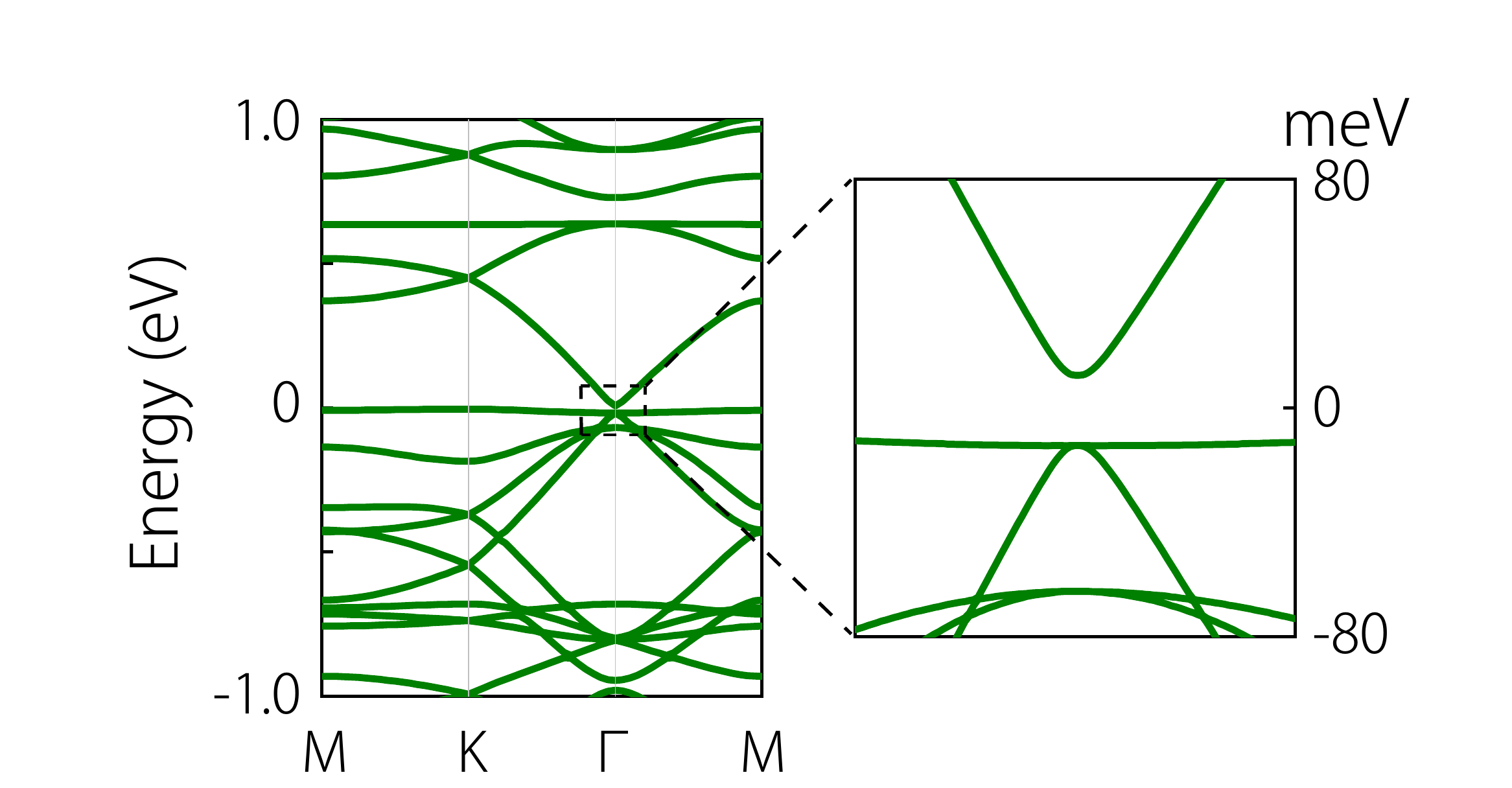}
\caption{Band structure of monolayer Co$_3$(HITP)$_2$ with SOC included. The right panel shows the enlarged view around band gap.
\label{fig3}}
\end{figure}

Next, we include SOC. The obtained band structure is shown in Fig.~\ref{fig3}. One can see that SOC effect is weak on the low-energy bands. This is understandable, because, as we mentioned, the low-energy states are mainly from N and C where SOC is weak; meanwhile, the relatively strong atomic SOC is from Co-$3d$ orbitals, but those are away from Fermi level.
After including SOC, the two spin channels are coupled together. Importantly, the combined symmetry $C_{2z}T$ is respected and we still have the algebra $(C_{2z}T)^2=1$~\footnote{With SOC, $T^2=-1$, and there is another minus sign from $C_{2z}^2=-1$ due to the spin-1/2 character.}, so the total real Chern number $\nu_R$ for the system remains well defined. However, now, $\nu_R$ can no longer be derived from $C_{2z}$ eigenvalues, since $C_{2z}$ is broken for the whole system by magnetism.

Here, we determine $\nu_R$ by a continuous deformation. We know that without SOC, the band structure (in Fig.~\ref{fig2}) is an insulator with $\nu_R=1$. Let us start from this band structure and gradually turn on SOC, e.g., by varying the SOC strength $\lambda$
from 0 to 1 with 1 representing the state of fully turned on. This can be easily done in first-principles calculations.
In this process, the defining symmetry $C_{2z}T$ is always preserved, so $\nu_R$ remains well defined as long as the system is an insulator. By monitoring this evolution, we find that the band gap keeps open in the whole process, meaning that the band structure with SOC can be continuously deformed to that without SOC. Thus, they must share the same $\nu_R=1$.
This analysis proves that 2D Co$_3$(HITP)$_2$ remains to be a MRCI in the presence of SOC.

\begin{figure}
\includegraphics[width=8.8cm]{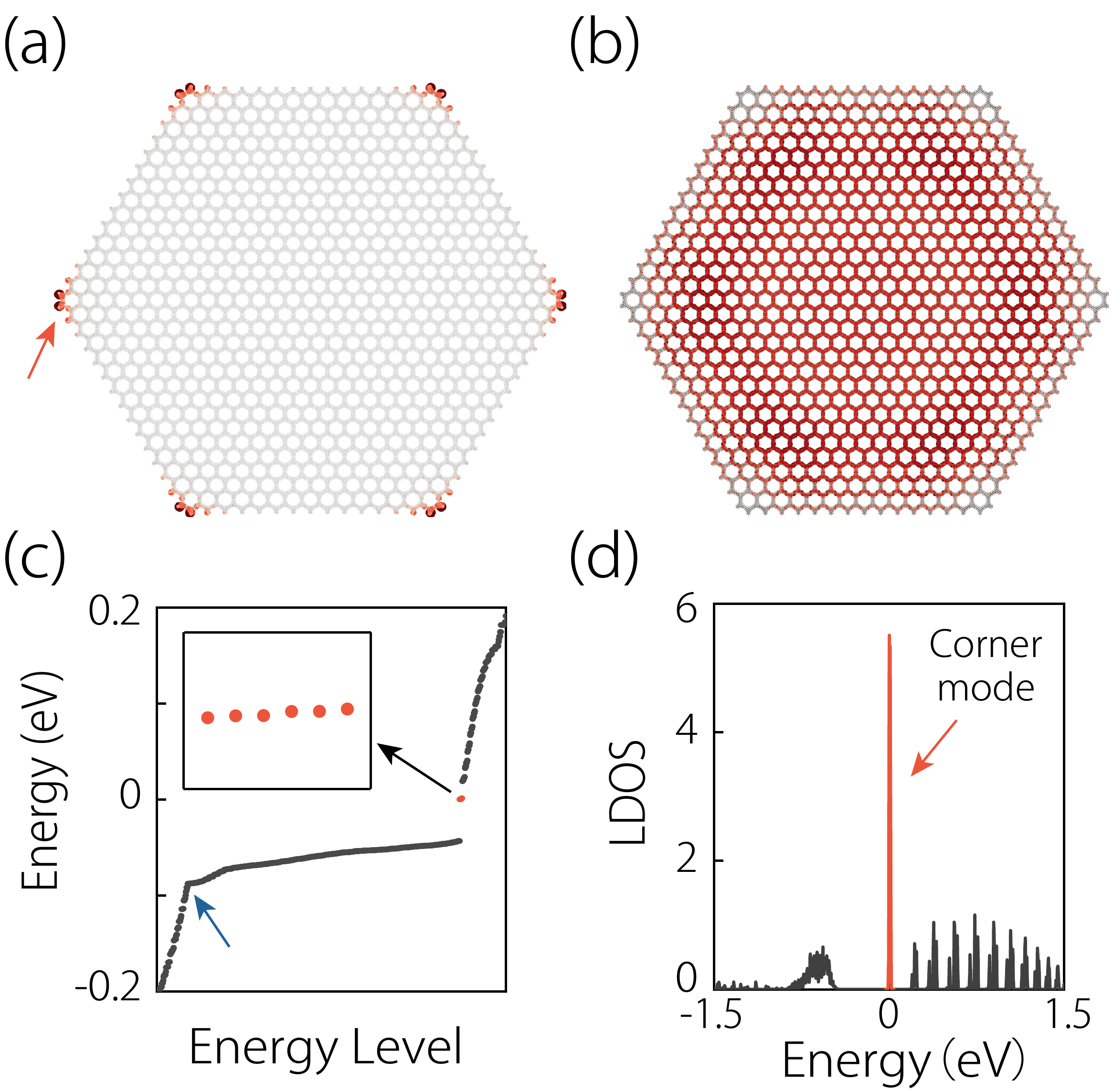}
\caption{(a) Spatial distribution of the corner modes (the red dots in (c)) for a Co$_3$(HITP)$_2$ disk. (b) Distribution of a typical bulk mode marked by blue arrow in (c). (c) shows the energy spectrum for the hexagonal shaped disk. (d) The \emph{local} density of states at one corner of the disk. The sharp peak corresponds to the corner zero-mode.
\label{fig4}}
\end{figure}

\emph{{\color{blue} Corner zero-modes.}} The bulk topology of a MRCI will generically manifest as topological zero-modes located at corners of the material. For a sample geometry that preserves $C_{2z}T$ symmetry, such modes appear in pairs connected by $C_{2z}T$~\cite{PhysRevB.104.085205}. To demonstrate this feature in Co$_3$(HITP)$_2$, we perform calculation for a hexagonal shaped disk as shown in Fig.~\ref{fig4}(a), which preserves $C_{2z}T$ and is the typical sample geometry obtained in experiment.

The calculated energy spectrum for the disk is plotted in Fig.~\ref{fig4}(c). One observes the existence of six zero-modes, i.e., modes at Fermi level in the bulk band gap. The tiny energy splitting among the six modes is due to finite-size effect, which diminishes with the increase of sample size.
Analysis of their wave function distribution confirms that they are localized at the six corners of the disk (see Fig.~\ref{fig4}(a)), so they are indeed corner zero-modes for MRCI. We also plot distribution of a representative bulk mode (marked by the blue arrow in Fig.~\ref{fig4}(c)), as shown in Fig.~\ref{fig4}(b). One can easily see the contrast. In Fig.~\ref{fig4}(d), we plot the \emph{local} density of states at one corner of the disk (taking the mode weight within one unit cell at the corner). One observes the sharp peak in bulk gap, corresponding to the corner zero-modes.

It should be noted that different from nonmagnetic real Chern insulators, such as the graphynes~\cite{PhysRevLett.123.256402,PhysRevB.104.085205}, the corner modes in
monolayer Co$_3$(HITP)$_2$  are strongly spin polarized. This can be readily understood from our analysis above.
Without SOC, these modes are fully spin polarized, as they belong to the spin-up channel with $\nu_R^\uparrow=1$.
After including SOC, they become not fully polarized, but since SOC effect is weak on the low-energy states, the polarization is still close to 100\%.
Such spin-polarized corner zero-modes can be readily detected in spin-resolved STS experiment~\cite{RevModPhys.81.1495}.

\emph{{\color{blue} Topological phase transition.}} 2D materials can usually sustain much larger strains ($\sim 10\%$) than 3D materials.
Experimental techniques for applying controlled strains on 2D materials are also well developed~\cite{kim2009large,https://doi.org/10.1002/adma.202205714}.
Noting that the band gap in 2D Co$_3$(HITP)$_2$ is small, we naturally expect that a small strain may close the gap and drive a topological/quantum phase transition.

From Fig.~\ref{fig3}, we observe that the two valence bands appear to form a double degeneracy at $\Gamma$. Symmetry analysis shows that
this degeneracy is exact in the absence of SOC, which belongs to the $E_{2u}$ irreducible representation of $D_{6h}$ group. It is no longer exact after including SOC, but since SOC effect is very weak for low-energy states (the energy splitting here is tiny $< 1$ meV), for the analysis here, we may safely neglect SOC effect.

\begin{figure}
\includegraphics[width=8.8cm]{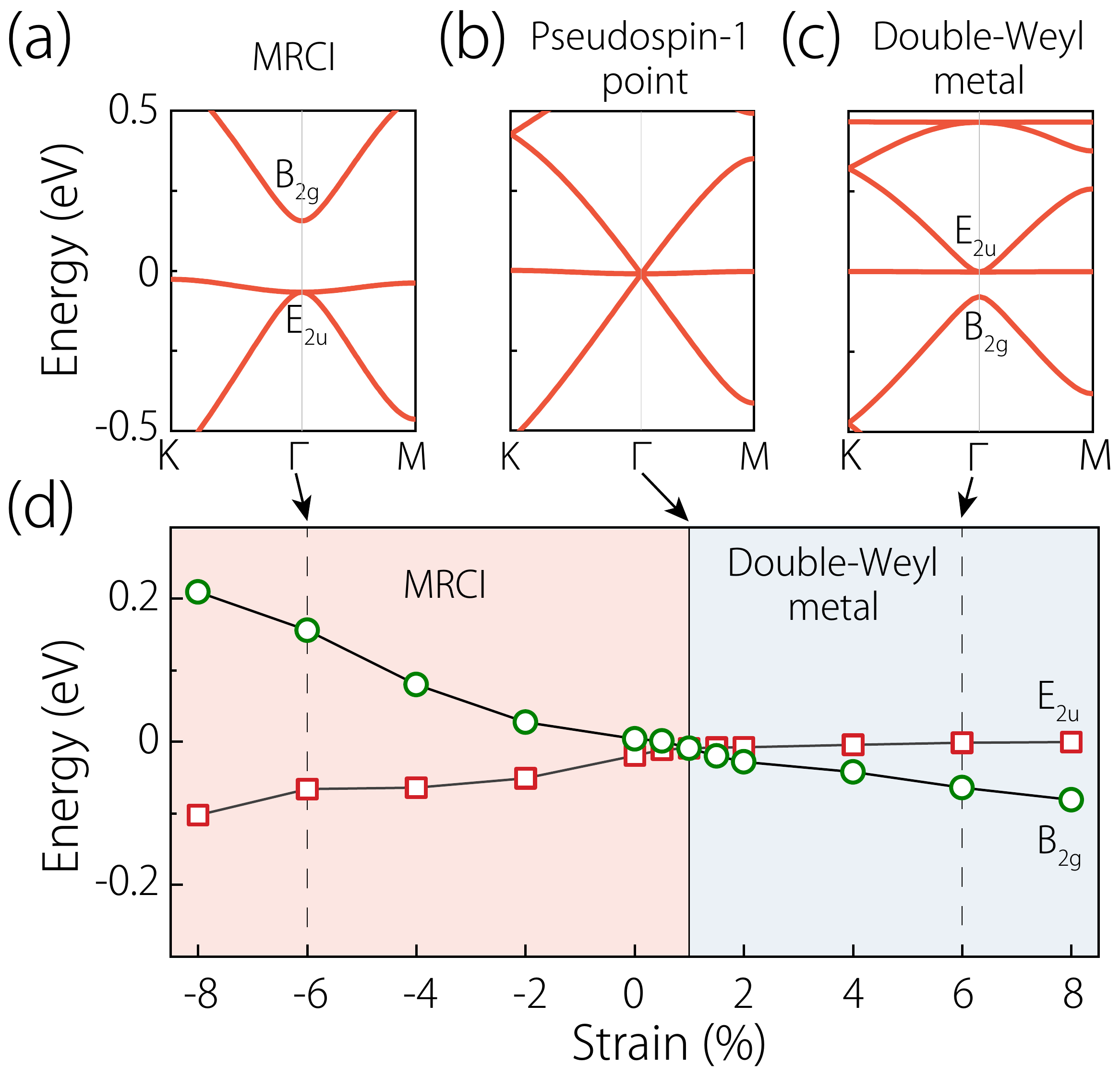}
\caption{Band structure of monolayer Co$_3$(HITP)$_2$ under (a) 6$\%$ compressive strain, (b) 1$\%$ tensile strain, and (c) 6$\%$ tensile strain. (d) shows the phase diagram of 2D Co$_3$(HITP)$_2$ under strain. The vertical axis shows variation of energies for $E_{2u}$ and $B_{2g}$ states.
\label{fig5}}
\end{figure}

In Fig.~\ref{fig5}(a-c), we plot the band structures for 2D Co$_3$(HITP)$_2$ under three representative biaxial strains. From the calculated strain-stress curves~\cite{refSM}, we find that monolayer Co$_3$(HITP)$_2$ can sustain at least 12\% biaxial strain.
Under compressive strain (Fig.~\ref{fig5}(a)), the system remains in the MRCI phase and its band gap will increase (Fig.~\ref{fig5}(d)).
The gap can be increased to 182 meV at $-6\%$ strain. Under tensile strain, the gap shrinks and closes at around 1\% strain (Fig.~\ref{fig5}(b)). Interestingly, at the critical point, the conduction band edge, which belong to $B_{2g}$ representation touches the $E_{2u}$ doublet, and they form a pseudospin-1 type dispersion, consisting of a Dirac cone plus a flat band. Further increasing strain,
$E_{2u}$ and $B_{2g}$ will switch order in energy. Note that due to band filling, the Fermi level should pass the $E_{2u}$ doublet, so the system transitions from MRCI into a metallic state. At the $E_{2u}$ degeneracy, the two bands touch quadratically. Such a degeneracy point is known as double Weyl point and was previously reported only in nonmagnetic 2D materials, such as blue phosphorene oxide~\cite{doi:10.1021/acs.nanolett.6b03208}, Mg$_2$C~\cite{PhysRevMaterials.2.104003}, and Na$_2$O~\cite{https://doi.org/10.1002/advs.201901939}. In comparison, the double Weyl point here has distinct features: (1) it is strongly spin-polarized; and (2) the lower band is very flat.
In Fig.~\ref{fig5}(d), we summarize these results in a phase diagram.

\emph{{\color{blue} Discussion.}} We have revealed 2D  Co$_3$(HITP)$_2$ as a concrete material candidate for the long-sought MRCI. Distinct from previously reported (nonmagnetic) real Chern insulators, the system here has intrinsic $T$ breaking,
topology robust under SOC, and spin-polarized corner zero-modes. Experimentally, the corner modes can be probed by STS. For example, one can compare the STS spectra measured at center and at corner of the disk, then the peak due to corner modes (as in Fig.~\ref{fig4}(d)) can be distinguished. The topological corner modes are robust against perturbations, which has been extensively addressed in previous studies~\cite{PhysRevLett.123.256402,PhysRevB.104.085205,PhysRevLett.119.246401,PhysRevLett.119.246402}. Especially since these modes are exponentially localized at the corners, they are insensitive to perturbations such as defects and impurities away from corners. Impurities at the corner may have a large influence, so they should be carefully minimized in experimental fabrication.

There exist a large class of MOF materials with transition metal ions. Besides Co$_3$(HITP)$_2$, we find similar physics in Mn$_3$(HITP)$_2$ and Fe$_3$(HITP)$_2$, which have also been synthesized in experiment~\cite{https://doi.org/10.1002/admt.202000941,doi:10.1021/acs.jpcc.0c08140}. As discussed in~\cite{refSM}, these two materials are also FM in ground state but the easy axis is out-of-plane. Their ground-state band structures are similar to that in Fig.~\ref{fig5}(c), belonging to the magnetic double-Weyl metal phase.

Finally, we mention that the flat band in the current system may offer a good platform for studying correlation and quantum effects. In the metallic phase in Fig.~\ref{fig5}(c), the flat band has a very small bandwidth $\sim $ 2 meV, implying strongly enhanced correlation effect under controlled hole doping. The existence of another quadratic touching band makes it a kind of singular flat band~\cite{rhim2020quantum}. Recent theory proposed that such a system offers opportunity to probe quantum metric via measuring the anomalous Landau level spectrum~\cite{rhim2020quantum}. Notably, unlike many previous setups where the flat band is spin degenerate, the band here is spin polarized.
This avoids the complication from Zeeman splitting, which is beneficial for probing the Landau spectrum.

\bibliographystyle{apsrev4-1}
\bibliography{ref}

\begin{thebibliography}{55}%
\makeatletter
\providecommand \@ifxundefined [1]{%
 \@ifx{#1\undefined}
}%
\providecommand \@ifnum [1]{%
 \ifnum #1\expandafter \@firstoftwo
 \else \expandafter \@secondoftwo
 \fi
}%
\providecommand \@ifx [1]{%
 \ifx #1\expandafter \@firstoftwo
 \else \expandafter \@secondoftwo
 \fi
}%
\providecommand \natexlab [1]{#1}%
\providecommand \enquote  [1]{``#1''}%
\providecommand \bibnamefont  [1]{#1}%
\providecommand \bibfnamefont [1]{#1}%
\providecommand \citenamefont [1]{#1}%
\providecommand \href@noop [0]{\@secondoftwo}%
\providecommand \href [0]{\begingroup \@sanitize@url \@href}%
\providecommand \@href[1]{\@@startlink{#1}\@@href}%
\providecommand \@@href[1]{\endgroup#1\@@endlink}%
\providecommand \@sanitize@url [0]{\catcode `\\12\catcode `\$12\catcode
  `\&12\catcode `\#12\catcode `\^12\catcode `\_12\catcode `\%12\relax}%
\providecommand \@@startlink[1]{}%
\providecommand \@@endlink[0]{}%
\providecommand \url  [0]{\begingroup\@sanitize@url \@url }%
\providecommand \@url [1]{\endgroup\@href {#1}{\urlprefix }}%
\providecommand \urlprefix  [0]{URL }%
\providecommand \Eprint [0]{\href }%
\providecommand \doibase [0]{http://dx.doi.org/}%
\providecommand \selectlanguage [0]{\@gobble}%
\providecommand \bibinfo  [0]{\@secondoftwo}%
\providecommand \bibfield  [0]{\@secondoftwo}%
\providecommand \translation [1]{[#1]}%
\providecommand \BibitemOpen [0]{}%
\providecommand \bibitemStop [0]{}%
\providecommand \bibitemNoStop [0]{.\EOS\space}%
\providecommand \EOS [0]{\spacefactor3000\relax}%
\providecommand \BibitemShut  [1]{\csname bibitem#1\endcsname}%
\let\auto@bib@innerbib\@empty
\bibitem [{\citenamefont {Hasan}\ and\ \citenamefont
  {Kane}(2010)}]{RevModPhys.82.3045}%
  \BibitemOpen
  \bibfield  {author} {\bibinfo {author} {\bibfnamefont {M.~Z.}\ \bibnamefont
  {Hasan}}\ and\ \bibinfo {author} {\bibfnamefont {C.~L.}\ \bibnamefont
  {Kane}},\ }\href {\doibase 10.1103/RevModPhys.82.3045} {\bibfield  {journal}
  {\bibinfo  {journal} {Rev. Mod. Phys.}\ }\textbf {\bibinfo {volume} {82}},\
  \bibinfo {pages} {3045} (\bibinfo {year} {2010})}\BibitemShut {NoStop}%
\bibitem [{\citenamefont {Qi}\ and\ \citenamefont
  {Zhang}(2011)}]{RevModPhys.83.1057}%
  \BibitemOpen
  \bibfield  {author} {\bibinfo {author} {\bibfnamefont {X.-L.}\ \bibnamefont
  {Qi}}\ and\ \bibinfo {author} {\bibfnamefont {S.-C.}\ \bibnamefont {Zhang}},\
  }\href {\doibase 10.1103/RevModPhys.83.1057} {\bibfield  {journal} {\bibinfo
  {journal} {Rev. Mod. Phys.}\ }\textbf {\bibinfo {volume} {83}},\ \bibinfo
  {pages} {1057} (\bibinfo {year} {2011})}\BibitemShut {NoStop}%
\bibitem [{\citenamefont {Shen}(2012)}]{shen2012topological}%
  \BibitemOpen
  \bibfield  {author} {\bibinfo {author} {\bibfnamefont {S.-Q.}\ \bibnamefont
  {Shen}},\ }\href@noop {} {\emph {\bibinfo {title} {Topological Insulators:
  Dirac Equation in Condensed Matters}}},\ Vol.\ \bibinfo {volume} {174}\
  (\bibinfo  {publisher} {Springer},\ \bibinfo {year} {2012})\BibitemShut
  {NoStop}%
\bibitem [{\citenamefont {Bansil}\ \emph {et~al.}(2016)\citenamefont {Bansil},
  \citenamefont {Lin},\ and\ \citenamefont {Das}}]{RevModPhys.88.021004}%
  \BibitemOpen
  \bibfield  {author} {\bibinfo {author} {\bibfnamefont {A.}~\bibnamefont
  {Bansil}}, \bibinfo {author} {\bibfnamefont {H.}~\bibnamefont {Lin}}, \ and\
  \bibinfo {author} {\bibfnamefont {T.}~\bibnamefont {Das}},\ }\href {\doibase
  10.1103/RevModPhys.88.021004} {\bibfield  {journal} {\bibinfo  {journal}
  {Rev. Mod. Phys.}\ }\textbf {\bibinfo {volume} {88}},\ \bibinfo {pages}
  {021004} (\bibinfo {year} {2016})}\BibitemShut {NoStop}%
\bibitem [{\citenamefont {Armitage}\ \emph {et~al.}(2018)\citenamefont
  {Armitage}, \citenamefont {Mele},\ and\ \citenamefont
  {Vishwanath}}]{RevModPhys.90.015001}%
  \BibitemOpen
  \bibfield  {author} {\bibinfo {author} {\bibfnamefont {N.~P.}\ \bibnamefont
  {Armitage}}, \bibinfo {author} {\bibfnamefont {E.~J.}\ \bibnamefont {Mele}},
  \ and\ \bibinfo {author} {\bibfnamefont {A.}~\bibnamefont {Vishwanath}},\
  }\href {\doibase 10.1103/RevModPhys.90.015001} {\bibfield  {journal}
  {\bibinfo  {journal} {Rev. Mod. Phys.}\ }\textbf {\bibinfo {volume} {90}},\
  \bibinfo {pages} {015001} (\bibinfo {year} {2018})}\BibitemShut {NoStop}%
\bibitem [{\citenamefont {Zhao}\ \emph {et~al.}(2016)\citenamefont {Zhao},
  \citenamefont {Schnyder},\ and\ \citenamefont
  {Wang}}]{PhysRevLett.116.156402}%
  \BibitemOpen
  \bibfield  {author} {\bibinfo {author} {\bibfnamefont {Y.~X.}\ \bibnamefont
  {Zhao}}, \bibinfo {author} {\bibfnamefont {A.~P.}\ \bibnamefont {Schnyder}},
  \ and\ \bibinfo {author} {\bibfnamefont {Z.~D.}\ \bibnamefont {Wang}},\
  }\href {\doibase 10.1103/PhysRevLett.116.156402} {\bibfield  {journal}
  {\bibinfo  {journal} {Phys. Rev. Lett.}\ }\textbf {\bibinfo {volume} {116}},\
  \bibinfo {pages} {156402} (\bibinfo {year} {2016})}\BibitemShut {NoStop}%
\bibitem [{\citenamefont {Zhao}\ and\ \citenamefont
  {Lu}(2017)}]{PhysRevLett.118.056401}%
  \BibitemOpen
  \bibfield  {author} {\bibinfo {author} {\bibfnamefont {Y.~X.}\ \bibnamefont
  {Zhao}}\ and\ \bibinfo {author} {\bibfnamefont {Y.}~\bibnamefont {Lu}},\
  }\href {\doibase 10.1103/PhysRevLett.118.056401} {\bibfield  {journal}
  {\bibinfo  {journal} {Phys. Rev. Lett.}\ }\textbf {\bibinfo {volume} {118}},\
  \bibinfo {pages} {056401} (\bibinfo {year} {2017})}\BibitemShut {NoStop}%
\bibitem [{\citenamefont {Bzdu\ifmmode~\check{s}\else \v{s}\fi{}ek}\ and\
  \citenamefont {Sigrist}(2017)}]{PhysRevB.96.155105}%
  \BibitemOpen
  \bibfield  {author} {\bibinfo {author} {\bibfnamefont {T.~c.~v.}\
  \bibnamefont {Bzdu\ifmmode~\check{s}\else \v{s}\fi{}ek}}\ and\ \bibinfo
  {author} {\bibfnamefont {M.}~\bibnamefont {Sigrist}},\ }\href {\doibase
  10.1103/PhysRevB.96.155105} {\bibfield  {journal} {\bibinfo  {journal} {Phys.
  Rev. B}\ }\textbf {\bibinfo {volume} {96}},\ \bibinfo {pages} {155105}
  (\bibinfo {year} {2017})}\BibitemShut {NoStop}%
\bibitem [{\citenamefont {Wu}\ \emph {et~al.}(2018)\citenamefont {Wu},
  \citenamefont {Liu}, \citenamefont {Li}, \citenamefont {Zhong}, \citenamefont
  {Yu}, \citenamefont {Sheng}, \citenamefont {Zhao},\ and\ \citenamefont
  {Yang}}]{PhysRevB.97.115125}%
  \BibitemOpen
  \bibfield  {author} {\bibinfo {author} {\bibfnamefont {W.}~\bibnamefont
  {Wu}}, \bibinfo {author} {\bibfnamefont {Y.}~\bibnamefont {Liu}}, \bibinfo
  {author} {\bibfnamefont {S.}~\bibnamefont {Li}}, \bibinfo {author}
  {\bibfnamefont {C.}~\bibnamefont {Zhong}}, \bibinfo {author} {\bibfnamefont
  {Z.-M.}\ \bibnamefont {Yu}}, \bibinfo {author} {\bibfnamefont {X.-L.}\
  \bibnamefont {Sheng}}, \bibinfo {author} {\bibfnamefont {Y.~X.}\ \bibnamefont
  {Zhao}}, \ and\ \bibinfo {author} {\bibfnamefont {S.~A.}\ \bibnamefont
  {Yang}},\ }\href {\doibase 10.1103/PhysRevB.97.115125} {\bibfield  {journal}
  {\bibinfo  {journal} {Phys. Rev. B}\ }\textbf {\bibinfo {volume} {97}},\
  \bibinfo {pages} {115125} (\bibinfo {year} {2018})}\BibitemShut {NoStop}%
\bibitem [{\citenamefont {Ahn}\ \emph {et~al.}(2018)\citenamefont {Ahn},
  \citenamefont {Kim}, \citenamefont {Kim},\ and\ \citenamefont
  {Yang}}]{PhysRevLett.121.106403}%
  \BibitemOpen
  \bibfield  {author} {\bibinfo {author} {\bibfnamefont {J.}~\bibnamefont
  {Ahn}}, \bibinfo {author} {\bibfnamefont {D.}~\bibnamefont {Kim}}, \bibinfo
  {author} {\bibfnamefont {Y.}~\bibnamefont {Kim}}, \ and\ \bibinfo {author}
  {\bibfnamefont {B.-J.}\ \bibnamefont {Yang}},\ }\href {\doibase
  10.1103/PhysRevLett.121.106403} {\bibfield  {journal} {\bibinfo  {journal}
  {Phys. Rev. Lett.}\ }\textbf {\bibinfo {volume} {121}},\ \bibinfo {pages}
  {106403} (\bibinfo {year} {2018})}\BibitemShut {NoStop}%
\bibitem [{\citenamefont {Wang}\ \emph
  {et~al.}(2020{\natexlab{a}})\citenamefont {Wang}, \citenamefont {Dai},
  \citenamefont {Shao}, \citenamefont {Yang},\ and\ \citenamefont
  {Zhao}}]{PhysRevLett.125.126403}%
  \BibitemOpen
  \bibfield  {author} {\bibinfo {author} {\bibfnamefont {K.}~\bibnamefont
  {Wang}}, \bibinfo {author} {\bibfnamefont {J.-X.}\ \bibnamefont {Dai}},
  \bibinfo {author} {\bibfnamefont {L.~B.}\ \bibnamefont {Shao}}, \bibinfo
  {author} {\bibfnamefont {S.~A.}\ \bibnamefont {Yang}}, \ and\ \bibinfo
  {author} {\bibfnamefont {Y.~X.}\ \bibnamefont {Zhao}},\ }\href {\doibase
  10.1103/PhysRevLett.125.126403} {\bibfield  {journal} {\bibinfo  {journal}
  {Phys. Rev. Lett.}\ }\textbf {\bibinfo {volume} {125}},\ \bibinfo {pages}
  {126403} (\bibinfo {year} {2020}{\natexlab{a}})}\BibitemShut {NoStop}%
\bibitem [{\citenamefont {Chen}\ \emph {et~al.}(2021)\citenamefont {Chen},
  \citenamefont {Wu}, \citenamefont {Yu}, \citenamefont {Chen}, \citenamefont
  {Zhao}, \citenamefont {Sheng},\ and\ \citenamefont
  {Yang}}]{PhysRevB.104.085205}%
  \BibitemOpen
  \bibfield  {author} {\bibinfo {author} {\bibfnamefont {C.}~\bibnamefont
  {Chen}}, \bibinfo {author} {\bibfnamefont {W.}~\bibnamefont {Wu}}, \bibinfo
  {author} {\bibfnamefont {Z.-M.}\ \bibnamefont {Yu}}, \bibinfo {author}
  {\bibfnamefont {Z.}~\bibnamefont {Chen}}, \bibinfo {author} {\bibfnamefont
  {Y.~X.}\ \bibnamefont {Zhao}}, \bibinfo {author} {\bibfnamefont {X.-L.}\
  \bibnamefont {Sheng}}, \ and\ \bibinfo {author} {\bibfnamefont {S.~A.}\
  \bibnamefont {Yang}},\ }\href {\doibase 10.1103/PhysRevB.104.085205}
  {\bibfield  {journal} {\bibinfo  {journal} {Phys. Rev. B}\ }\textbf {\bibinfo
  {volume} {104}},\ \bibinfo {pages} {085205} (\bibinfo {year}
  {2021})}\BibitemShut {NoStop}%
\bibitem [{\citenamefont {Wu}\ \emph {et~al.}(2019)\citenamefont {Wu},
  \citenamefont {Soluyanov},\ and\ \citenamefont
  {Bzdušek}}]{doi:10.1126/science.aau8740}%
  \BibitemOpen
  \bibfield  {author} {\bibinfo {author} {\bibfnamefont {Q.}~\bibnamefont
  {Wu}}, \bibinfo {author} {\bibfnamefont {A.~A.}\ \bibnamefont {Soluyanov}}, \
  and\ \bibinfo {author} {\bibfnamefont {T.}~\bibnamefont {Bzdušek}},\ }\href
  {\doibase 10.1126/science.aau8740} {\bibfield  {journal} {\bibinfo  {journal}
  {Science}\ }\textbf {\bibinfo {volume} {365}},\ \bibinfo {pages} {1273}
  (\bibinfo {year} {2019})}\BibitemShut {NoStop}%
\bibitem [{\citenamefont {Sheng}\ \emph {et~al.}(2019)\citenamefont {Sheng},
  \citenamefont {Chen}, \citenamefont {Liu}, \citenamefont {Chen},
  \citenamefont {Yu}, \citenamefont {Zhao},\ and\ \citenamefont
  {Yang}}]{PhysRevLett.123.256402}%
  \BibitemOpen
  \bibfield  {author} {\bibinfo {author} {\bibfnamefont {X.-L.}\ \bibnamefont
  {Sheng}}, \bibinfo {author} {\bibfnamefont {C.}~\bibnamefont {Chen}},
  \bibinfo {author} {\bibfnamefont {H.}~\bibnamefont {Liu}}, \bibinfo {author}
  {\bibfnamefont {Z.}~\bibnamefont {Chen}}, \bibinfo {author} {\bibfnamefont
  {Z.-M.}\ \bibnamefont {Yu}}, \bibinfo {author} {\bibfnamefont {Y.~X.}\
  \bibnamefont {Zhao}}, \ and\ \bibinfo {author} {\bibfnamefont {S.~A.}\
  \bibnamefont {Yang}},\ }\href {\doibase 10.1103/PhysRevLett.123.256402}
  {\bibfield  {journal} {\bibinfo  {journal} {Phys. Rev. Lett.}\ }\textbf
  {\bibinfo {volume} {123}},\ \bibinfo {pages} {256402} (\bibinfo {year}
  {2019})}\BibitemShut {NoStop}%
\bibitem [{\citenamefont {Lee}\ \emph {et~al.}(2020)\citenamefont {Lee},
  \citenamefont {Kim}, \citenamefont {Ahn},\ and\ \citenamefont
  {Yang}}]{lee2020two}%
  \BibitemOpen
  \bibfield  {author} {\bibinfo {author} {\bibfnamefont {E.}~\bibnamefont
  {Lee}}, \bibinfo {author} {\bibfnamefont {R.}~\bibnamefont {Kim}}, \bibinfo
  {author} {\bibfnamefont {J.}~\bibnamefont {Ahn}}, \ and\ \bibinfo {author}
  {\bibfnamefont {B.-J.}\ \bibnamefont {Yang}},\ }\href {\doibase
  10.1038/s41535-019-0206-8} {\bibfield  {journal} {\bibinfo  {journal} {npj
  Quantum Mater.}\ }\textbf {\bibinfo {volume} {5}},\ \bibinfo {pages} {1}
  (\bibinfo {year} {2020})}\BibitemShut {NoStop}%
\bibitem [{\citenamefont {Chen}\ \emph {et~al.}(2022)\citenamefont {Chen},
  \citenamefont {Zeng}, \citenamefont {Chen}, \citenamefont {Zhao},
  \citenamefont {Sheng},\ and\ \citenamefont {Yang}}]{PhysRevLett.128.026405}%
  \BibitemOpen
  \bibfield  {author} {\bibinfo {author} {\bibfnamefont {C.}~\bibnamefont
  {Chen}}, \bibinfo {author} {\bibfnamefont {X.-T.}\ \bibnamefont {Zeng}},
  \bibinfo {author} {\bibfnamefont {Z.}~\bibnamefont {Chen}}, \bibinfo {author}
  {\bibfnamefont {Y.~X.}\ \bibnamefont {Zhao}}, \bibinfo {author}
  {\bibfnamefont {X.-L.}\ \bibnamefont {Sheng}}, \ and\ \bibinfo {author}
  {\bibfnamefont {S.~A.}\ \bibnamefont {Yang}},\ }\href {\doibase
  10.1103/PhysRevLett.128.026405} {\bibfield  {journal} {\bibinfo  {journal}
  {Phys. Rev. Lett.}\ }\textbf {\bibinfo {volume} {128}},\ \bibinfo {pages}
  {026405} (\bibinfo {year} {2022})}\BibitemShut {NoStop}%
\bibitem [{\citenamefont {Zhu}\ \emph {et~al.}(2022)\citenamefont {Zhu},
  \citenamefont {Wu}, \citenamefont {Zhao}, \citenamefont {Chen}, \citenamefont
  {Wang}, \citenamefont {Sheng}, \citenamefont {Zhang}, \citenamefont {Zhao},\
  and\ \citenamefont {Yang}}]{PhysRevB.105.085123}%
  \BibitemOpen
  \bibfield  {author} {\bibinfo {author} {\bibfnamefont {J.}~\bibnamefont
  {Zhu}}, \bibinfo {author} {\bibfnamefont {W.}~\bibnamefont {Wu}}, \bibinfo
  {author} {\bibfnamefont {J.}~\bibnamefont {Zhao}}, \bibinfo {author}
  {\bibfnamefont {C.}~\bibnamefont {Chen}}, \bibinfo {author} {\bibfnamefont
  {Q.}~\bibnamefont {Wang}}, \bibinfo {author} {\bibfnamefont {X.-L.}\
  \bibnamefont {Sheng}}, \bibinfo {author} {\bibfnamefont {L.}~\bibnamefont
  {Zhang}}, \bibinfo {author} {\bibfnamefont {Y.~X.}\ \bibnamefont {Zhao}}, \
  and\ \bibinfo {author} {\bibfnamefont {S.~A.}\ \bibnamefont {Yang}},\ }\href
  {\doibase 10.1103/PhysRevB.105.085123} {\bibfield  {journal} {\bibinfo
  {journal} {Phys. Rev. B}\ }\textbf {\bibinfo {volume} {105}},\ \bibinfo
  {pages} {085123} (\bibinfo {year} {2022})}\BibitemShut {NoStop}%
\bibitem [{Note1()}]{Note1}%
  \BibitemOpen
  \bibinfo {note} {This is due to the algebra $(PT)^2=\pm 1$ for
  spinless/spinful cases. Nevertheless, with certain gauge flux distribution,
  the two cases may be switched. See Ref~\cite
  {PhysRevLett.126.196402}.}\BibitemShut {Stop}%
\bibitem [{\citenamefont {Ahn}\ and\ \citenamefont
  {Yang}(2019)}]{PhysRevB.99.235125}%
  \BibitemOpen
  \bibfield  {author} {\bibinfo {author} {\bibfnamefont {J.}~\bibnamefont
  {Ahn}}\ and\ \bibinfo {author} {\bibfnamefont {B.-J.}\ \bibnamefont {Yang}},\
  }\href {\doibase 10.1103/PhysRevB.99.235125} {\bibfield  {journal} {\bibinfo
  {journal} {Phys. Rev. B}\ }\textbf {\bibinfo {volume} {99}},\ \bibinfo
  {pages} {235125} (\bibinfo {year} {2019})}\BibitemShut {NoStop}%
\bibitem [{\citenamefont {Ahn}\ \emph {et~al.}(2019)\citenamefont {Ahn},
  \citenamefont {Park},\ and\ \citenamefont {Yang}}]{PhysRevX.9.021013}%
  \BibitemOpen
  \bibfield  {author} {\bibinfo {author} {\bibfnamefont {J.}~\bibnamefont
  {Ahn}}, \bibinfo {author} {\bibfnamefont {S.}~\bibnamefont {Park}}, \ and\
  \bibinfo {author} {\bibfnamefont {B.-J.}\ \bibnamefont {Yang}},\ }\href
  {\doibase 10.1103/PhysRevX.9.021013} {\bibfield  {journal} {\bibinfo
  {journal} {Phys. Rev. X}\ }\textbf {\bibinfo {volume} {9}},\ \bibinfo {pages}
  {021013} (\bibinfo {year} {2019})}\BibitemShut {NoStop}%
\bibitem [{\citenamefont {Li}\ \emph {et~al.}(1999)\citenamefont {Li},
  \citenamefont {Eddaoudi}, \citenamefont {O'Keeffe},\ and\ \citenamefont
  {Yaghi}}]{li1999design}%
  \BibitemOpen
  \bibfield  {author} {\bibinfo {author} {\bibfnamefont {H.}~\bibnamefont
  {Li}}, \bibinfo {author} {\bibfnamefont {M.}~\bibnamefont {Eddaoudi}},
  \bibinfo {author} {\bibfnamefont {M.}~\bibnamefont {O'Keeffe}}, \ and\
  \bibinfo {author} {\bibfnamefont {O.~M.}\ \bibnamefont {Yaghi}},\ }\href
  {\doibase 10.1038/46248} {\bibfield  {journal} {\bibinfo  {journal} {Nature}\
  }\textbf {\bibinfo {volume} {402}},\ \bibinfo {pages} {276} (\bibinfo {year}
  {1999})}\BibitemShut {NoStop}%
\bibitem [{\citenamefont {Furukawa}\ \emph {et~al.}(2013)\citenamefont
  {Furukawa}, \citenamefont {Cordova}, \citenamefont {O'Keeffe},\ and\
  \citenamefont {Yaghi}}]{WOS:000323652300033}%
  \BibitemOpen
  \bibfield  {author} {\bibinfo {author} {\bibfnamefont {H.}~\bibnamefont
  {Furukawa}}, \bibinfo {author} {\bibfnamefont {K.~E.}\ \bibnamefont
  {Cordova}}, \bibinfo {author} {\bibfnamefont {M.}~\bibnamefont {O'Keeffe}}, \
  and\ \bibinfo {author} {\bibfnamefont {O.~M.}\ \bibnamefont {Yaghi}},\ }\href
  {\doibase 10.1126/science.1230444} {\bibfield  {journal} {\bibinfo  {journal}
  {Science}\ }\textbf {\bibinfo {volume} {341}},\ \bibinfo {pages} {1230444}
  (\bibinfo {year} {2013})}\BibitemShut {NoStop}%
\bibitem [{\citenamefont {Pedersen}\ \emph {et~al.}(2018)\citenamefont
  {Pedersen}, \citenamefont {Perlepe}, \citenamefont {Aubrey}, \citenamefont
  {Woodruff}, \citenamefont {Reyes-Lillo}, \citenamefont {Reinholdt},
  \citenamefont {Voigt}, \citenamefont {Li}, \citenamefont {Borup},
  \citenamefont {Rouzieres}, \citenamefont {Samohvalov}, \citenamefont
  {Wilhelm}, \citenamefont {Rogalev}, \citenamefont {Neaton}, \citenamefont
  {Long},\ and\ \citenamefont {Clerac}}]{WOS:000445129300013}%
  \BibitemOpen
  \bibfield  {author} {\bibinfo {author} {\bibfnamefont {K.~S.}\ \bibnamefont
  {Pedersen}}, \bibinfo {author} {\bibfnamefont {P.}~\bibnamefont {Perlepe}},
  \bibinfo {author} {\bibfnamefont {M.~L.}\ \bibnamefont {Aubrey}}, \bibinfo
  {author} {\bibfnamefont {D.~N.}\ \bibnamefont {Woodruff}}, \bibinfo {author}
  {\bibfnamefont {S.~E.}\ \bibnamefont {Reyes-Lillo}}, \bibinfo {author}
  {\bibfnamefont {A.}~\bibnamefont {Reinholdt}}, \bibinfo {author}
  {\bibfnamefont {L.}~\bibnamefont {Voigt}}, \bibinfo {author} {\bibfnamefont
  {Z.}~\bibnamefont {Li}}, \bibinfo {author} {\bibfnamefont {K.}~\bibnamefont
  {Borup}}, \bibinfo {author} {\bibfnamefont {M.}~\bibnamefont {Rouzieres}},
  \bibinfo {author} {\bibfnamefont {D.}~\bibnamefont {Samohvalov}}, \bibinfo
  {author} {\bibfnamefont {F.}~\bibnamefont {Wilhelm}}, \bibinfo {author}
  {\bibfnamefont {A.}~\bibnamefont {Rogalev}}, \bibinfo {author} {\bibfnamefont
  {J.~B.}\ \bibnamefont {Neaton}}, \bibinfo {author} {\bibfnamefont {J.~R.}\
  \bibnamefont {Long}}, \ and\ \bibinfo {author} {\bibfnamefont
  {R.}~\bibnamefont {Clerac}},\ }\href {\doibase 10.1038/s41557-018-0107-7}
  {\bibfield  {journal} {\bibinfo  {journal} {Nat. Chem.}\ }\textbf {\bibinfo
  {volume} {10}},\ \bibinfo {pages} {1056} (\bibinfo {year}
  {2018})}\BibitemShut {NoStop}%
\bibitem [{\citenamefont {Dong}\ \emph {et~al.}(2018)\citenamefont {Dong},
  \citenamefont {Zhang}, \citenamefont {Tranca}, \citenamefont {Zhou},
  \citenamefont {Wang}, \citenamefont {Adler}, \citenamefont {Liao},
  \citenamefont {Liu}, \citenamefont {Sun}, \citenamefont {Shi}, \citenamefont
  {Zhang}, \citenamefont {Zschech}, \citenamefont {Mannsfeld}, \citenamefont
  {Felser},\ and\ \citenamefont {Feng}}]{WOS:000437677800014}%
  \BibitemOpen
  \bibfield  {author} {\bibinfo {author} {\bibfnamefont {R.}~\bibnamefont
  {Dong}}, \bibinfo {author} {\bibfnamefont {Z.}~\bibnamefont {Zhang}},
  \bibinfo {author} {\bibfnamefont {D.~C.}\ \bibnamefont {Tranca}}, \bibinfo
  {author} {\bibfnamefont {S.}~\bibnamefont {Zhou}}, \bibinfo {author}
  {\bibfnamefont {M.}~\bibnamefont {Wang}}, \bibinfo {author} {\bibfnamefont
  {P.}~\bibnamefont {Adler}}, \bibinfo {author} {\bibfnamefont
  {Z.}~\bibnamefont {Liao}}, \bibinfo {author} {\bibfnamefont {F.}~\bibnamefont
  {Liu}}, \bibinfo {author} {\bibfnamefont {Y.}~\bibnamefont {Sun}}, \bibinfo
  {author} {\bibfnamefont {W.}~\bibnamefont {Shi}}, \bibinfo {author}
  {\bibfnamefont {Z.}~\bibnamefont {Zhang}}, \bibinfo {author} {\bibfnamefont
  {E.}~\bibnamefont {Zschech}}, \bibinfo {author} {\bibfnamefont {S.~C.~B.}\
  \bibnamefont {Mannsfeld}}, \bibinfo {author} {\bibfnamefont {C.}~\bibnamefont
  {Felser}}, \ and\ \bibinfo {author} {\bibfnamefont {X.}~\bibnamefont
  {Feng}},\ }\href {\doibase 10.1038/s41467-018-05141-4} {\bibfield  {journal}
  {\bibinfo  {journal} {Nat. Commun.}\ }\textbf {\bibinfo {volume} {9}},\
  \bibinfo {pages} {2637} (\bibinfo {year} {2018})}\BibitemShut {NoStop}%
\bibitem [{\citenamefont {Yuan}\ \emph {et~al.}(2019)\citenamefont {Yuan},
  \citenamefont {Song}, \citenamefont {Zhu}, \citenamefont {Li}, \citenamefont
  {Han}, \citenamefont {Zheng}, \citenamefont {Li}, \citenamefont {Zhang},\
  and\ \citenamefont {Hu}}]{https://doi.org/10.1002/smll.201804845}%
  \BibitemOpen
  \bibfield  {author} {\bibinfo {author} {\bibfnamefont {K.}~\bibnamefont
  {Yuan}}, \bibinfo {author} {\bibfnamefont {T.}~\bibnamefont {Song}}, \bibinfo
  {author} {\bibfnamefont {X.}~\bibnamefont {Zhu}}, \bibinfo {author}
  {\bibfnamefont {B.}~\bibnamefont {Li}}, \bibinfo {author} {\bibfnamefont
  {B.}~\bibnamefont {Han}}, \bibinfo {author} {\bibfnamefont {L.}~\bibnamefont
  {Zheng}}, \bibinfo {author} {\bibfnamefont {J.}~\bibnamefont {Li}}, \bibinfo
  {author} {\bibfnamefont {X.}~\bibnamefont {Zhang}}, \ and\ \bibinfo {author}
  {\bibfnamefont {W.}~\bibnamefont {Hu}},\ }\href {\doibase
  https://doi.org/10.1002/smll.201804845} {\bibfield  {journal} {\bibinfo
  {journal} {Small}\ }\textbf {\bibinfo {volume} {15}},\ \bibinfo {pages}
  {1804845} (\bibinfo {year} {2019})}\BibitemShut {NoStop}%
\bibitem [{\citenamefont {Xia}\ \emph {et~al.}(2015)\citenamefont {Xia},
  \citenamefont {Mahmood}, \citenamefont {Zou},\ and\ \citenamefont
  {Xu}}]{C5EE00762C}%
  \BibitemOpen
  \bibfield  {author} {\bibinfo {author} {\bibfnamefont {W.}~\bibnamefont
  {Xia}}, \bibinfo {author} {\bibfnamefont {A.}~\bibnamefont {Mahmood}},
  \bibinfo {author} {\bibfnamefont {R.}~\bibnamefont {Zou}}, \ and\ \bibinfo
  {author} {\bibfnamefont {Q.}~\bibnamefont {Xu}},\ }\href {\doibase
  10.1039/C5EE00762C} {\bibfield  {journal} {\bibinfo  {journal} {Energy
  Environ. Sci.}\ }\textbf {\bibinfo {volume} {8}},\ \bibinfo {pages} {1837}
  (\bibinfo {year} {2015})}\BibitemShut {NoStop}%
\bibitem [{\citenamefont {Murray}\ \emph {et~al.}(2009)\citenamefont {Murray},
  \citenamefont {Dincă},\ and\ \citenamefont {Long}}]{B802256A}%
  \BibitemOpen
  \bibfield  {author} {\bibinfo {author} {\bibfnamefont {L.~J.}\ \bibnamefont
  {Murray}}, \bibinfo {author} {\bibfnamefont {M.}~\bibnamefont {Dincă}}, \
  and\ \bibinfo {author} {\bibfnamefont {J.~R.}\ \bibnamefont {Long}},\ }\href
  {\doibase 10.1039/B802256A} {\bibfield  {journal} {\bibinfo  {journal} {Chem.
  Soc. Rev.}\ }\textbf {\bibinfo {volume} {38}},\ \bibinfo {pages} {1294}
  (\bibinfo {year} {2009})}\BibitemShut {NoStop}%
\bibitem [{\citenamefont {Zhang}\ \emph {et~al.}(2019)\citenamefont {Zhang},
  \citenamefont {Chen}, \citenamefont {Zhong}, \citenamefont {Zhang},
  \citenamefont {Zhang}, \citenamefont {Zhou},\ and\ \citenamefont
  {Bu}}]{WOS:000606751900002}%
  \BibitemOpen
  \bibfield  {author} {\bibinfo {author} {\bibfnamefont {X.}~\bibnamefont
  {Zhang}}, \bibinfo {author} {\bibfnamefont {A.}~\bibnamefont {Chen}},
  \bibinfo {author} {\bibfnamefont {M.}~\bibnamefont {Zhong}}, \bibinfo
  {author} {\bibfnamefont {Z.}~\bibnamefont {Zhang}}, \bibinfo {author}
  {\bibfnamefont {X.}~\bibnamefont {Zhang}}, \bibinfo {author} {\bibfnamefont
  {Z.}~\bibnamefont {Zhou}}, \ and\ \bibinfo {author} {\bibfnamefont {X.-H.}\
  \bibnamefont {Bu}},\ }\href {\doibase 10.1007/s41918-018-0024-x} {\bibfield
  {journal} {\bibinfo  {journal} {Electrochem. Energ. Rev.}\ }\textbf {\bibinfo
  {volume} {2}},\ \bibinfo {pages} {29} (\bibinfo {year} {2019})}\BibitemShut
  {NoStop}%
\bibitem [{\citenamefont {Lee}\ \emph {et~al.}(2009)\citenamefont {Lee},
  \citenamefont {Farha}, \citenamefont {Roberts}, \citenamefont {Scheidt},
  \citenamefont {Nguyen},\ and\ \citenamefont {Hupp}}]{B807080F}%
  \BibitemOpen
  \bibfield  {author} {\bibinfo {author} {\bibfnamefont {J.}~\bibnamefont
  {Lee}}, \bibinfo {author} {\bibfnamefont {O.~K.}\ \bibnamefont {Farha}},
  \bibinfo {author} {\bibfnamefont {J.}~\bibnamefont {Roberts}}, \bibinfo
  {author} {\bibfnamefont {K.~A.}\ \bibnamefont {Scheidt}}, \bibinfo {author}
  {\bibfnamefont {S.~T.}\ \bibnamefont {Nguyen}}, \ and\ \bibinfo {author}
  {\bibfnamefont {J.~T.}\ \bibnamefont {Hupp}},\ }\href {\doibase
  10.1039/B807080F} {\bibfield  {journal} {\bibinfo  {journal} {Chem. Soc.
  Rev.}\ }\textbf {\bibinfo {volume} {38}},\ \bibinfo {pages} {1450} (\bibinfo
  {year} {2009})}\BibitemShut {NoStop}%
\bibitem [{\citenamefont {Liu}\ \emph {et~al.}(2014)\citenamefont {Liu},
  \citenamefont {Chen}, \citenamefont {Cui}, \citenamefont {Zhang},
  \citenamefont {Zhang},\ and\ \citenamefont {Su}}]{C4CS00094C}%
  \BibitemOpen
  \bibfield  {author} {\bibinfo {author} {\bibfnamefont {J.}~\bibnamefont
  {Liu}}, \bibinfo {author} {\bibfnamefont {L.}~\bibnamefont {Chen}}, \bibinfo
  {author} {\bibfnamefont {H.}~\bibnamefont {Cui}}, \bibinfo {author}
  {\bibfnamefont {J.}~\bibnamefont {Zhang}}, \bibinfo {author} {\bibfnamefont
  {L.}~\bibnamefont {Zhang}}, \ and\ \bibinfo {author} {\bibfnamefont {C.-Y.}\
  \bibnamefont {Su}},\ }\href {\doibase 10.1039/C4CS00094C} {\bibfield
  {journal} {\bibinfo  {journal} {Chem. Soc. Rev.}\ }\textbf {\bibinfo {volume}
  {43}},\ \bibinfo {pages} {6011} (\bibinfo {year} {2014})}\BibitemShut
  {NoStop}%
\bibitem [{\citenamefont {Jiao}\ \emph {et~al.}(2017)\citenamefont {Jiao},
  \citenamefont {Wang}, \citenamefont {Jiang},\ and\ \citenamefont
  {Xu}}]{https://doi.org/10.1002/adma.201703663}%
  \BibitemOpen
  \bibfield  {author} {\bibinfo {author} {\bibfnamefont {L.}~\bibnamefont
  {Jiao}}, \bibinfo {author} {\bibfnamefont {Y.}~\bibnamefont {Wang}}, \bibinfo
  {author} {\bibfnamefont {H.-L.}\ \bibnamefont {Jiang}}, \ and\ \bibinfo
  {author} {\bibfnamefont {Q.}~\bibnamefont {Xu}},\ }\href {\doibase
  https://doi.org/10.1002/adma.201703663} {\bibfield  {journal} {\bibinfo
  {journal} {Adv. Mater.}\ }\textbf {\bibinfo {volume} {30}},\ \bibinfo {pages}
  {1703663} (\bibinfo {year} {2017})}\BibitemShut {NoStop}%
\bibitem [{\citenamefont {Kreno}\ \emph {et~al.}(2012)\citenamefont {Kreno},
  \citenamefont {Leong}, \citenamefont {Farha}, \citenamefont {Allendorf},
  \citenamefont {Van~Duyne},\ and\ \citenamefont
  {Hupp}}]{doi:10.1021/cr200324t}%
  \BibitemOpen
  \bibfield  {author} {\bibinfo {author} {\bibfnamefont {L.~E.}\ \bibnamefont
  {Kreno}}, \bibinfo {author} {\bibfnamefont {K.}~\bibnamefont {Leong}},
  \bibinfo {author} {\bibfnamefont {O.~K.}\ \bibnamefont {Farha}}, \bibinfo
  {author} {\bibfnamefont {M.}~\bibnamefont {Allendorf}}, \bibinfo {author}
  {\bibfnamefont {R.~P.}\ \bibnamefont {Van~Duyne}}, \ and\ \bibinfo {author}
  {\bibfnamefont {J.~T.}\ \bibnamefont {Hupp}},\ }\href {\doibase
  10.1021/cr200324t} {\bibfield  {journal} {\bibinfo  {journal} {Chem. Rev.}\
  }\textbf {\bibinfo {volume} {112}},\ \bibinfo {pages} {1105} (\bibinfo {year}
  {2012})}\BibitemShut {NoStop}%
\bibitem [{\citenamefont {Li}\ \emph {et~al.}(2020)\citenamefont {Li},
  \citenamefont {Zhao}, \citenamefont {Zang},\ and\ \citenamefont
  {Li}}]{C9CS00778D}%
  \BibitemOpen
  \bibfield  {author} {\bibinfo {author} {\bibfnamefont {H.-Y.}\ \bibnamefont
  {Li}}, \bibinfo {author} {\bibfnamefont {S.-N.}\ \bibnamefont {Zhao}},
  \bibinfo {author} {\bibfnamefont {S.-Q.}\ \bibnamefont {Zang}}, \ and\
  \bibinfo {author} {\bibfnamefont {J.}~\bibnamefont {Li}},\ }\href {\doibase
  10.1039/C9CS00778D} {\bibfield  {journal} {\bibinfo  {journal} {Chem. Soc.
  Rev.}\ }\textbf {\bibinfo {volume} {49}},\ \bibinfo {pages} {6364} (\bibinfo
  {year} {2020})}\BibitemShut {NoStop}%
\bibitem [{\citenamefont {Wang}\ \emph {et~al.}(2013)\citenamefont {Wang},
  \citenamefont {Liu},\ and\ \citenamefont {Liu}}]{PhysRevLett.110.196801}%
  \BibitemOpen
  \bibfield  {author} {\bibinfo {author} {\bibfnamefont {Z.~F.}\ \bibnamefont
  {Wang}}, \bibinfo {author} {\bibfnamefont {Z.}~\bibnamefont {Liu}}, \ and\
  \bibinfo {author} {\bibfnamefont {F.}~\bibnamefont {Liu}},\ }\href {\doibase
  10.1103/PhysRevLett.110.196801} {\bibfield  {journal} {\bibinfo  {journal}
  {Phys. Rev. Lett.}\ }\textbf {\bibinfo {volume} {110}},\ \bibinfo {pages}
  {196801} (\bibinfo {year} {2013})}\BibitemShut {NoStop}%
\bibitem [{\citenamefont {Yamada}\ \emph {et~al.}(2016)\citenamefont {Yamada},
  \citenamefont {Soejima}, \citenamefont {Tsuji}, \citenamefont {Hirai},
  \citenamefont {Dinc\ifmmode~\u{a}\else \u{a}\fi{}},\ and\ \citenamefont
  {Aoki}}]{PhysRevB.94.081102}%
  \BibitemOpen
  \bibfield  {author} {\bibinfo {author} {\bibfnamefont {M.~G.}\ \bibnamefont
  {Yamada}}, \bibinfo {author} {\bibfnamefont {T.}~\bibnamefont {Soejima}},
  \bibinfo {author} {\bibfnamefont {N.}~\bibnamefont {Tsuji}}, \bibinfo
  {author} {\bibfnamefont {D.}~\bibnamefont {Hirai}}, \bibinfo {author}
  {\bibfnamefont {M.}~\bibnamefont {Dinc\ifmmode~\u{a}\else \u{a}\fi{}}}, \
  and\ \bibinfo {author} {\bibfnamefont {H.}~\bibnamefont {Aoki}},\ }\href
  {\doibase 10.1103/PhysRevB.94.081102} {\bibfield  {journal} {\bibinfo
  {journal} {Phys. Rev. B}\ }\textbf {\bibinfo {volume} {94}},\ \bibinfo
  {pages} {081102} (\bibinfo {year} {2016})}\BibitemShut {NoStop}%
\bibitem [{\citenamefont {Jiang}\ \emph {et~al.}(2020)\citenamefont {Jiang},
  \citenamefont {Zhang}, \citenamefont {Wang}, \citenamefont {Liu},\ and\
  \citenamefont {Low}}]{doi:10.1021/acs.nanolett.9b05242}%
  \BibitemOpen
  \bibfield  {author} {\bibinfo {author} {\bibfnamefont {W.}~\bibnamefont
  {Jiang}}, \bibinfo {author} {\bibfnamefont {S.}~\bibnamefont {Zhang}},
  \bibinfo {author} {\bibfnamefont {Z.}~\bibnamefont {Wang}}, \bibinfo {author}
  {\bibfnamefont {F.}~\bibnamefont {Liu}}, \ and\ \bibinfo {author}
  {\bibfnamefont {T.}~\bibnamefont {Low}},\ }\href {\doibase
  10.1021/acs.nanolett.9b05242} {\bibfield  {journal} {\bibinfo  {journal}
  {Nano Lett.}\ }\textbf {\bibinfo {volume} {20}},\ \bibinfo {pages} {1959}
  (\bibinfo {year} {2020})}\BibitemShut {NoStop}%
\bibitem [{\citenamefont {Jiang}\ \emph {et~al.}(2021)\citenamefont {Jiang},
  \citenamefont {Ni},\ and\ \citenamefont
  {Liu}}]{doi:10.1021/acs.accounts.0c00652}%
  \BibitemOpen
  \bibfield  {author} {\bibinfo {author} {\bibfnamefont {W.}~\bibnamefont
  {Jiang}}, \bibinfo {author} {\bibfnamefont {X.}~\bibnamefont {Ni}}, \ and\
  \bibinfo {author} {\bibfnamefont {F.}~\bibnamefont {Liu}},\ }\href {\doibase
  10.1021/acs.accounts.0c00652} {\bibfield  {journal} {\bibinfo  {journal}
  {Acc. Chem. Res.}\ }\textbf {\bibinfo {volume} {54}},\ \bibinfo {pages} {416}
  (\bibinfo {year} {2021})}\BibitemShut {NoStop}%
\bibitem [{\citenamefont {Chen}\ \emph {et~al.}(2020)\citenamefont {Chen},
  \citenamefont {Dou}, \citenamefont {Yang}, \citenamefont {Sun}, \citenamefont
  {Libretto}, \citenamefont {Skorupskii}, \citenamefont {Miller},\ and\
  \citenamefont {Dincă}}]{doi:10.1021/jacs.0c04458}%
  \BibitemOpen
  \bibfield  {author} {\bibinfo {author} {\bibfnamefont {T.}~\bibnamefont
  {Chen}}, \bibinfo {author} {\bibfnamefont {J.-H.}\ \bibnamefont {Dou}},
  \bibinfo {author} {\bibfnamefont {L.}~\bibnamefont {Yang}}, \bibinfo {author}
  {\bibfnamefont {C.}~\bibnamefont {Sun}}, \bibinfo {author} {\bibfnamefont
  {N.~J.}\ \bibnamefont {Libretto}}, \bibinfo {author} {\bibfnamefont
  {G.}~\bibnamefont {Skorupskii}}, \bibinfo {author} {\bibfnamefont {J.~T.}\
  \bibnamefont {Miller}}, \ and\ \bibinfo {author} {\bibfnamefont
  {M.}~\bibnamefont {Dincă}},\ }\href {\doibase 10.1021/jacs.0c04458}
  {\bibfield  {journal} {\bibinfo  {journal} {J. Am. Chem. Soc.}\ }\textbf
  {\bibinfo {volume} {142}},\ \bibinfo {pages} {12367} (\bibinfo {year}
  {2020})}\BibitemShut {NoStop}%
\bibitem [{\citenamefont {Iqbal}\ \emph {et~al.}(2021)\citenamefont {Iqbal},
  \citenamefont {Sultan}, \citenamefont {Hussain}, \citenamefont {Hamza},
  \citenamefont {Tariq}, \citenamefont {Akbar}, \citenamefont {Ma},\ and\
  \citenamefont {Zhi}}]{https://doi.org/10.1002/admt.202000941}%
  \BibitemOpen
  \bibfield  {author} {\bibinfo {author} {\bibfnamefont {R.}~\bibnamefont
  {Iqbal}}, \bibinfo {author} {\bibfnamefont {M.~Q.}\ \bibnamefont {Sultan}},
  \bibinfo {author} {\bibfnamefont {S.}~\bibnamefont {Hussain}}, \bibinfo
  {author} {\bibfnamefont {M.}~\bibnamefont {Hamza}}, \bibinfo {author}
  {\bibfnamefont {A.}~\bibnamefont {Tariq}}, \bibinfo {author} {\bibfnamefont
  {M.~B.}\ \bibnamefont {Akbar}}, \bibinfo {author} {\bibfnamefont
  {Y.}~\bibnamefont {Ma}}, \ and\ \bibinfo {author} {\bibfnamefont
  {L.}~\bibnamefont {Zhi}},\ }\href {\doibase
  https://doi.org/10.1002/admt.202000941} {\bibfield  {journal} {\bibinfo
  {journal} {Adv. Mater. Technol.}\ }\textbf {\bibinfo {volume} {6}},\ \bibinfo
  {pages} {2000941} (\bibinfo {year} {2021})}\BibitemShut {NoStop}%
\bibitem [{\citenamefont {Gao}\ \emph {et~al.}(2020)\citenamefont {Gao},
  \citenamefont {Gao}, \citenamefont {Hua}, \citenamefont {Liu}, \citenamefont
  {Huang},\ and\ \citenamefont {Lin}}]{doi:10.1021/acs.jpcc.0c08140}%
  \BibitemOpen
  \bibfield  {author} {\bibinfo {author} {\bibfnamefont {Z.}~\bibnamefont
  {Gao}}, \bibinfo {author} {\bibfnamefont {Y.}~\bibnamefont {Gao}}, \bibinfo
  {author} {\bibfnamefont {M.}~\bibnamefont {Hua}}, \bibinfo {author}
  {\bibfnamefont {J.}~\bibnamefont {Liu}}, \bibinfo {author} {\bibfnamefont
  {L.}~\bibnamefont {Huang}}, \ and\ \bibinfo {author} {\bibfnamefont
  {N.}~\bibnamefont {Lin}},\ }\href {\doibase 10.1021/acs.jpcc.0c08140}
  {\bibfield  {journal} {\bibinfo  {journal} {J. Phys. Chem. C}\ }\textbf
  {\bibinfo {volume} {124}},\ \bibinfo {pages} {27017} (\bibinfo {year}
  {2020})}\BibitemShut {NoStop}%
\bibitem [{ref()}]{refSM}%
  \BibitemOpen
  \href@noop {} {}\bibinfo {note} {See Supplemental Material for the considered
  magnetic configurations, the parity of all occupied bands of monolayer
  $\mathrm{Co}_{3}\mathrm{HITP}_{2}$, the stress-strain relationship of
  monolayer $\mathrm{Co}_{3}\mathrm{HITP}_{2}$ under biaxial strain and the
  electronic band structure of monolayer $\mathrm{Fe}_{3}\mathrm{HITP}_{2}$ and
  $\mathrm{Mn}_{3}\mathrm{HITP}_{2}$}\BibitemShut {NoStop}%
\bibitem [{\citenamefont {Wang}\ \emph
  {et~al.}(2020{\natexlab{b}})\citenamefont {Wang}, \citenamefont {Fan},
  \citenamefont {Qi}, \citenamefont {Li},\ and\ \citenamefont
  {Zhao}}]{doi:10.1021/acs.jpcc.0c01143}%
  \BibitemOpen
  \bibfield  {author} {\bibinfo {author} {\bibfnamefont {J.}~\bibnamefont
  {Wang}}, \bibinfo {author} {\bibfnamefont {Y.}~\bibnamefont {Fan}}, \bibinfo
  {author} {\bibfnamefont {S.}~\bibnamefont {Qi}}, \bibinfo {author}
  {\bibfnamefont {W.}~\bibnamefont {Li}}, \ and\ \bibinfo {author}
  {\bibfnamefont {M.}~\bibnamefont {Zhao}},\ }\href {\doibase
  10.1021/acs.jpcc.0c01143} {\bibfield  {journal} {\bibinfo  {journal} {J.
  Phys. Chem. C}\ }\textbf {\bibinfo {volume} {124}},\ \bibinfo {pages} {9350}
  (\bibinfo {year} {2020}{\natexlab{b}})}\BibitemShut {NoStop}%
\bibitem [{\citenamefont {Kang}\ and\ \citenamefont {Yu}(2022)}]{D2CP02612K}%
  \BibitemOpen
  \bibfield  {author} {\bibinfo {author} {\bibfnamefont {S.}~\bibnamefont
  {Kang}}\ and\ \bibinfo {author} {\bibfnamefont {J.}~\bibnamefont {Yu}},\
  }\href {\doibase 10.1039/D2CP02612K} {\bibfield  {journal} {\bibinfo
  {journal} {Phys. Chem. Chem. Phys.}\ }\textbf {\bibinfo {volume} {24}},\
  \bibinfo {pages} {22168} (\bibinfo {year} {2022})}\BibitemShut {NoStop}%
\bibitem [{\citenamefont {Vanderbilt}(2018)}]{vanderbilt2018berry}%
  \BibitemOpen
  \bibfield  {author} {\bibinfo {author} {\bibfnamefont {D.}~\bibnamefont
  {Vanderbilt}},\ }\href@noop {} {\emph {\bibinfo {title} {Berry phases in
  electronic structure theory: electric polarization, orbital magnetization and
  topological insulators}}}\ (\bibinfo  {publisher} {Cambridge University
  Press},\ \bibinfo {year} {2018})\BibitemShut {NoStop}%
\bibitem [{Note2()}]{Note2}%
  \BibitemOpen
  \bibinfo {note} {With SOC, $T^2=-1$, and there is another minus sign from
  $C_{2z}^2=-1$ due to the spin-1/2 character.}\BibitemShut {Stop}%
\bibitem [{\citenamefont {Wiesendanger}(2009)}]{RevModPhys.81.1495}%
  \BibitemOpen
  \bibfield  {author} {\bibinfo {author} {\bibfnamefont {R.}~\bibnamefont
  {Wiesendanger}},\ }\href {\doibase 10.1103/RevModPhys.81.1495} {\bibfield
  {journal} {\bibinfo  {journal} {Rev. Mod. Phys.}\ }\textbf {\bibinfo {volume}
  {81}},\ \bibinfo {pages} {1495} (\bibinfo {year} {2009})}\BibitemShut
  {NoStop}%
\bibitem [{\citenamefont {Kim}\ \emph {et~al.}(2009)\citenamefont {Kim},
  \citenamefont {Zhao}, \citenamefont {Jang}, \citenamefont {Lee},
  \citenamefont {Kim}, \citenamefont {Kim}, \citenamefont {Ahn}, \citenamefont
  {Kim}, \citenamefont {Choi},\ and\ \citenamefont {Hong}}]{kim2009large}%
  \BibitemOpen
  \bibfield  {author} {\bibinfo {author} {\bibfnamefont {K.~S.}\ \bibnamefont
  {Kim}}, \bibinfo {author} {\bibfnamefont {Y.}~\bibnamefont {Zhao}}, \bibinfo
  {author} {\bibfnamefont {H.}~\bibnamefont {Jang}}, \bibinfo {author}
  {\bibfnamefont {S.~Y.}\ \bibnamefont {Lee}}, \bibinfo {author} {\bibfnamefont
  {J.~M.}\ \bibnamefont {Kim}}, \bibinfo {author} {\bibfnamefont {K.~S.}\
  \bibnamefont {Kim}}, \bibinfo {author} {\bibfnamefont {J.-H.}\ \bibnamefont
  {Ahn}}, \bibinfo {author} {\bibfnamefont {P.}~\bibnamefont {Kim}}, \bibinfo
  {author} {\bibfnamefont {J.-Y.}\ \bibnamefont {Choi}}, \ and\ \bibinfo
  {author} {\bibfnamefont {B.~H.}\ \bibnamefont {Hong}},\ }\href {\doibase
  https://doi.org/10.1038/nature07719} {\bibfield  {journal} {\bibinfo
  {journal} {Nature}\ }\textbf {\bibinfo {volume} {457}},\ \bibinfo {pages}
  {706} (\bibinfo {year} {2009})}\BibitemShut {NoStop}%
\bibitem [{\citenamefont {Qi}\ \emph {et~al.}(2023)\citenamefont {Qi},
  \citenamefont {Sadi}, \citenamefont {Hu}, \citenamefont {Zheng},
  \citenamefont {Wu}, \citenamefont {Jiang},\ and\ \citenamefont
  {Chen}}]{https://doi.org/10.1002/adma.202205714}%
  \BibitemOpen
  \bibfield  {author} {\bibinfo {author} {\bibfnamefont {Y.}~\bibnamefont
  {Qi}}, \bibinfo {author} {\bibfnamefont {M.~A.}\ \bibnamefont {Sadi}},
  \bibinfo {author} {\bibfnamefont {D.}~\bibnamefont {Hu}}, \bibinfo {author}
  {\bibfnamefont {M.}~\bibnamefont {Zheng}}, \bibinfo {author} {\bibfnamefont
  {Z.}~\bibnamefont {Wu}}, \bibinfo {author} {\bibfnamefont {Y.}~\bibnamefont
  {Jiang}}, \ and\ \bibinfo {author} {\bibfnamefont {Y.~P.}\ \bibnamefont
  {Chen}},\ }\href {\doibase https://doi.org/10.1002/adma.202205714} {\bibfield
   {journal} {\bibinfo  {journal} {Adv. Mater.}\ }\textbf {\bibinfo {volume}
  {n/a}},\ \bibinfo {pages} {2205714} (\bibinfo {year} {2023})}\BibitemShut
  {NoStop}%
\bibitem [{\citenamefont {Zhu}\ \emph {et~al.}(2016)\citenamefont {Zhu},
  \citenamefont {Wang}, \citenamefont {Guan}, \citenamefont {Liu},
  \citenamefont {Zhang}, \citenamefont {Chen},\ and\ \citenamefont
  {Yang}}]{doi:10.1021/acs.nanolett.6b03208}%
  \BibitemOpen
  \bibfield  {author} {\bibinfo {author} {\bibfnamefont {L.}~\bibnamefont
  {Zhu}}, \bibinfo {author} {\bibfnamefont {S.-S.}\ \bibnamefont {Wang}},
  \bibinfo {author} {\bibfnamefont {S.}~\bibnamefont {Guan}}, \bibinfo {author}
  {\bibfnamefont {Y.}~\bibnamefont {Liu}}, \bibinfo {author} {\bibfnamefont
  {T.}~\bibnamefont {Zhang}}, \bibinfo {author} {\bibfnamefont
  {G.}~\bibnamefont {Chen}}, \ and\ \bibinfo {author} {\bibfnamefont {S.~A.}\
  \bibnamefont {Yang}},\ }\href {\doibase 10.1021/acs.nanolett.6b03208}
  {\bibfield  {journal} {\bibinfo  {journal} {Nano Lett.}\ }\textbf {\bibinfo
  {volume} {16}},\ \bibinfo {pages} {6548} (\bibinfo {year}
  {2016})}\BibitemShut {NoStop}%
\bibitem [{\citenamefont {Wang}\ \emph {et~al.}(2018)\citenamefont {Wang},
  \citenamefont {Liu}, \citenamefont {Yu}, \citenamefont {Sheng}, \citenamefont
  {Zhu}, \citenamefont {Guan},\ and\ \citenamefont
  {Yang}}]{PhysRevMaterials.2.104003}%
  \BibitemOpen
  \bibfield  {author} {\bibinfo {author} {\bibfnamefont {S.-S.}\ \bibnamefont
  {Wang}}, \bibinfo {author} {\bibfnamefont {Y.}~\bibnamefont {Liu}}, \bibinfo
  {author} {\bibfnamefont {Z.-M.}\ \bibnamefont {Yu}}, \bibinfo {author}
  {\bibfnamefont {X.-L.}\ \bibnamefont {Sheng}}, \bibinfo {author}
  {\bibfnamefont {L.}~\bibnamefont {Zhu}}, \bibinfo {author} {\bibfnamefont
  {S.}~\bibnamefont {Guan}}, \ and\ \bibinfo {author} {\bibfnamefont {S.~A.}\
  \bibnamefont {Yang}},\ }\href {\doibase 10.1103/PhysRevMaterials.2.104003}
  {\bibfield  {journal} {\bibinfo  {journal} {Phys. Rev. Mater.}\ }\textbf
  {\bibinfo {volume} {2}},\ \bibinfo {pages} {104003} (\bibinfo {year}
  {2018})}\BibitemShut {NoStop}%
\bibitem [{\citenamefont {Hua}\ \emph {et~al.}(2020)\citenamefont {Hua},
  \citenamefont {Li}, \citenamefont {Xu}, \citenamefont {Zheng}, \citenamefont
  {Yang},\ and\ \citenamefont {Lu}}]{https://doi.org/10.1002/advs.201901939}%
  \BibitemOpen
  \bibfield  {author} {\bibinfo {author} {\bibfnamefont {C.}~\bibnamefont
  {Hua}}, \bibinfo {author} {\bibfnamefont {S.}~\bibnamefont {Li}}, \bibinfo
  {author} {\bibfnamefont {Z.-A.}\ \bibnamefont {Xu}}, \bibinfo {author}
  {\bibfnamefont {Y.}~\bibnamefont {Zheng}}, \bibinfo {author} {\bibfnamefont
  {S.~A.}\ \bibnamefont {Yang}}, \ and\ \bibinfo {author} {\bibfnamefont
  {Y.}~\bibnamefont {Lu}},\ }\href {\doibase
  https://doi.org/10.1002/advs.201901939} {\bibfield  {journal} {\bibinfo
  {journal} {Adv. Sci.}\ }\textbf {\bibinfo {volume} {7}},\ \bibinfo {pages}
  {1901939} (\bibinfo {year} {2020})}\BibitemShut {NoStop}%
\bibitem [{\citenamefont {Langbehn}\ \emph {et~al.}(2017)\citenamefont
  {Langbehn}, \citenamefont {Peng}, \citenamefont {Trifunovic}, \citenamefont
  {von Oppen},\ and\ \citenamefont {Brouwer}}]{PhysRevLett.119.246401}%
  \BibitemOpen
  \bibfield  {author} {\bibinfo {author} {\bibfnamefont {J.}~\bibnamefont
  {Langbehn}}, \bibinfo {author} {\bibfnamefont {Y.}~\bibnamefont {Peng}},
  \bibinfo {author} {\bibfnamefont {L.}~\bibnamefont {Trifunovic}}, \bibinfo
  {author} {\bibfnamefont {F.}~\bibnamefont {von Oppen}}, \ and\ \bibinfo
  {author} {\bibfnamefont {P.~W.}\ \bibnamefont {Brouwer}},\ }\href {\doibase
  10.1103/PhysRevLett.119.246401} {\bibfield  {journal} {\bibinfo  {journal}
  {Phys. Rev. Lett.}\ }\textbf {\bibinfo {volume} {119}},\ \bibinfo {pages}
  {246401} (\bibinfo {year} {2017})}\BibitemShut {NoStop}%
\bibitem [{\citenamefont {Song}\ \emph {et~al.}(2017)\citenamefont {Song},
  \citenamefont {Fang},\ and\ \citenamefont {Fang}}]{PhysRevLett.119.246402}%
  \BibitemOpen
  \bibfield  {author} {\bibinfo {author} {\bibfnamefont {Z.}~\bibnamefont
  {Song}}, \bibinfo {author} {\bibfnamefont {Z.}~\bibnamefont {Fang}}, \ and\
  \bibinfo {author} {\bibfnamefont {C.}~\bibnamefont {Fang}},\ }\href {\doibase
  10.1103/PhysRevLett.119.246402} {\bibfield  {journal} {\bibinfo  {journal}
  {Phys. Rev. Lett.}\ }\textbf {\bibinfo {volume} {119}},\ \bibinfo {pages}
  {246402} (\bibinfo {year} {2017})}\BibitemShut {NoStop}%
\bibitem [{\citenamefont {Rhim}\ \emph {et~al.}(2020)\citenamefont {Rhim},
  \citenamefont {Kim},\ and\ \citenamefont {Yang}}]{rhim2020quantum}%
  \BibitemOpen
  \bibfield  {author} {\bibinfo {author} {\bibfnamefont {J.-W.}\ \bibnamefont
  {Rhim}}, \bibinfo {author} {\bibfnamefont {K.}~\bibnamefont {Kim}}, \ and\
  \bibinfo {author} {\bibfnamefont {B.-J.}\ \bibnamefont {Yang}},\ }\href
  {\doibase 10.1038/s41586-020-2540-1} {\bibfield  {journal} {\bibinfo
  {journal} {Nature}\ }\textbf {\bibinfo {volume} {584}},\ \bibinfo {pages}
  {59} (\bibinfo {year} {2020})}\BibitemShut {NoStop}%
\bibitem [{\citenamefont {Zhao}\ \emph {et~al.}(2021)\citenamefont {Zhao},
  \citenamefont {Chen}, \citenamefont {Sheng},\ and\ \citenamefont
  {Yang}}]{PhysRevLett.126.196402}%
  \BibitemOpen
  \bibfield  {author} {\bibinfo {author} {\bibfnamefont {Y.~X.}\ \bibnamefont
  {Zhao}}, \bibinfo {author} {\bibfnamefont {C.}~\bibnamefont {Chen}}, \bibinfo
  {author} {\bibfnamefont {X.-L.}\ \bibnamefont {Sheng}}, \ and\ \bibinfo
  {author} {\bibfnamefont {S.~A.}\ \bibnamefont {Yang}},\ }\href {\doibase
  10.1103/PhysRevLett.126.196402} {\bibfield  {journal} {\bibinfo  {journal}
  {Phys. Rev. Lett.}\ }\textbf {\bibinfo {volume} {126}},\ \bibinfo {pages}
  {196402} (\bibinfo {year} {2021})}\BibitemShut {NoStop}%
\end{thebibliography}%

\end{document}